\documentclass{aa}  

\usepackage{graphicx}
\usepackage{xcolor}
\usepackage{txfonts}
\usepackage{tablefootnote}
\usepackage{threeparttable}
\usepackage{booktabs}
\usepackage[normalem]{ulem}

\usepackage[T1]{fontenc}

\newcommand{\galpak}{GalPaK$^{\rm 3D}$}

\newcommand{\OII}{\hbox{{\rm O}{\sc \,ii}}}

\newcommand{\OIII}{\hbox{{\rm O}{\sc \,iii}}}

\newcommand{\MgII}{\hbox{{\rm Mg}{\sc \,ii}}}

\newcommand{\HI}{\hbox{{\rm H}{\sc \,i}}}

\newcommand{\lla}{\hbox{$\lambda\lambda 2796,2803$}}
\newcommand{\Ha}{\hbox{{\rm H}$\alpha$}}

\newcommand{\kms}{\hbox{km~s$^{-1}$}}

\newcommand{\msun}{\hbox{M$_{\odot}$}}

\newcommand{\REW}{\hbox{$W_{\rm r}^{2796}$}}
\begin{document}

   \title{MusE GAs FLOw and Wind (MEGAFLOW) XIII. Cool  gas traced by \MgII\ around isolated galaxies}


   \author{Maxime Cherrey \inst{1}, 
          Nicolas F. Bouch\'e \inst{1},
          Johannes Zabl \inst{1,2},
          Ilane Schroetter \inst{3},
          Martin Wendt \inst{4},
          Ivanna Langan \inst{5},
          Joop Schaye \inst{6},
          Lutz Wisotzki \inst{7},
          Yucheng Guo \inst{1},
          Ismael Pessa \inst{7}.
          }

   \institute{Univ Lyon1, Ens de Lyon, CNRS, Centre de Recherche Astrophysique de Lyon (CRAL) UMR5574, F-69230 Saint- Genis-Laval, France\\
         \and
        Institute for Computational Astrophysics and Department of Astronomy \&\ Physics, Saint Mary's University, 923 Robie Street, Halifax, Nova Scotia, B3H 3C3, Canada 
        \and
        Institut de Recherche en Astrophysique et Plan\'etologie (IRAP), Universit\'e de Toulouse, CNRS, UPS, F-31400 Toulouse, France
        \and
        Institut f\"{u}r Physik und Astronomie, Universit\"{a}t Potsdam, Karl-Liebknecht-Str. 24/25, 14476 Potsdam, Germany 
        \and
        Centro de Astrobiolog\'ia (CAB), CSIC-INTA, Carretera de Ajalvir km 4, Torrej\'on de Ardoz, 28850, Madrid, Spain
        \and
        Leiden Observatory, Leiden University, P.O.Box 9513, NL-2300 AA Leiden, The Netherlands
        \and
        Leibniz-Institut for Astrophysik Potsdam (AIP), An der Sternwarte 16, 14482 Potsdam, Germany
        }

   \date{Received June 18 2024; revised Dec 5 2024}

 
  \abstract
  
   {The circumgalactic medium (CGM) is a key component needed to understand the physical processes governing the flows of gas around galaxies. Quantifying its evolution and its dependence on galaxy properties is particularly important for our understanding of accretion and feedback mechanisms.}
    {We select a volume-selected sample of 66 {\it isolated} star-forming  galaxies (SFGs) at $0.4< z <1.5$ with $\log(M_\star/\msun)> 9$ from the MusE GAs FLOw and Wind (MEGAFLOW) survey. Using \MgII\lla{} absorptions in background quasars, we measure the covering fraction $f_c$ and quantify
    how the cool gas profile depends on galaxy properties (such as star-formation rate (SFR), stellar mass ($M_\star$) or azimuthal angle relative to the line of sight) and how these dependencies evolve with redshift.}
    {The \MgII{} covering fraction of isolated galaxies is a strong function of impact parameter, and is steeper than previously reported.
    The impact parameter $b_{50}$ at which $f_c = $50\%  is $b_{50}=50\pm7$~kpc ($65\pm7$~kpc) for \REW$>$0.5~\AA\ (\REW$>$0.1~\AA), respectively. It is weakly correlated with SFR ($\propto$ SFR$^{0.08\pm0.09}$) and  decreases with cosmic time ($\propto (1+z)^{0.8 \pm 0.7}$), contrary to the expectation of increasingly larger halos with time. The covering fraction is also higher along the minor axis   than along the major axis  at the $\approx 2 \sigma$ level. }
    {The CGM traced by \MgII{} is similar across the isolated galaxy population. Indeed, among the isolated galaxies with an impact parameter below 55 kpc, all have associated \MgII{} absorption with \REW$>$0.3\AA,
    resulting in a steep covering fraction $f_c(b)$.
    }
    {}


\titlerunning{MEGAFLOW XIII: cool gas around isolated galaxies}
\authorrunning{Cherrey et al.}
   
   \maketitle

%

\section{Introduction}

The Circum-Galactic Medium (CGM) is a crucial element in understanding the formation and evolution of galaxies as it constitutes an important reservoir of baryonic matter. It is a complex multi-phase medium governed by multiple physical processes \citep{Faucher-Giguere_2023} Observations of the diffuse gas in emission is challenging and require to focus on resonant lines such as Ly$\alpha$ \citep[e.g ][]{Wisotzki_2018} and/or to use stacking techniques and deep observations \citep[e.g][]{Zhang_2016, Guo_2023}. Consequently, the CGM is often studied in absorption by searching for specific lines in bright background sources such as quasars. In particular, the \MgII\ absorption doublet \lla\ is known to be a good tracer of \HI\ because of it's low ionization potential. It is used to study the cool phase at $\approx 10^4$K \citep{Bergeron_1986, Bergeron_1991}. 
Theory and observations seem to indicate that this phase is clumpy, with a volume filling factor $\lesssim 1$\%  and sub-kpc clouds \citep{Chelouche_2009, Fukugita_2017, Faerman_2023, Hummels_2023, Afruni_2023}. The \MgII\ doublet has the advantage to be accessible in the optical from ground-based telescopes at intermediate redshifts $0.3 \lesssim z \lesssim 1.8$, where the cosmic star formation is at its peak. 

Several studies have shown how \MgII\ absorption equivalent width decreases as a function of the impact parameter $b$ to the line of sight \citep{Steidel_1995, Bordoloi_2011, Kacprzak_2013, Nielsen_2013, Rubin_2014, Schroetter_2019, Lan_2020, Dutta_2020, Dutta_2021, Huang_2021, Lundgren_2021}. The \MgII-$b$ anti-correlation extends to $\approx$ 50 to 150 kpc, with a large amount of scatter. This scatter can be explained first, by the natural variations among the galaxy populations, and second, by the orientation and inclination of the galaxy relative to the line of sight, as outflows are preferentially ejected perpendicular to the galactic plane and accretion is more likely to happen in the galactic plane \citep{Chen_2010_Na, Kacprzak_2012, Ho_2017, Ho_2019, Zabl_2019, Schroetter_2019,  Peroux_2020_simulations, Beckett_2021, Nateghi_2024a, Nateghi_2023b, Das_2024}.

An additional source of scatter can also be the presence of neighbours close to the line of sight due to the natural spatial correlation of galaxies. Indeed, in quasar absorption studies multiple galaxies are often detected at the absorption redshift within impact parameters of 100 kpc, making the galaxy/absorption pairing difficult. The solution is often to assume that the absorption is caused by the closest or the brightest galaxy. This approach is convenient but may be inaccurate. In addition, some neighbours may remain undetected due to completeness issues or finite field of view. The presence of neighbours (detected or undetected) is problematic if we aim to measure the cool CGM profile because the "superposed" contributions \citep{Bordoloi_2011, Nielsen_2018} from individual galaxies increase the absorption strength.

In order to avoid this source of confusion we select from the MusE GAs FLOw and Wind (MEGAFLOW) survey (presented in section \ref{sec:data}) a mass complete sample of isolated galaxies (defined in section \ref{sec:isolation_criteria}), for which we determine the morphology and kinematics using the forward Bayesian modelling algorithm \galpak (section \ref{sec:galaxy_properties}). We quantify in details the cool gas profile and the \MgII\ covering fraction for this sample, and investigate how they are impacted by galaxy properties (section \ref{sec:MgII_halos}).

Throughout, we use a standard flat $\Lambda$CDM cosmology with $H_0 = 69.3$~km~s\textsuperscript{-1} Mpc\textsuperscript{-1}, $\Omega_M = 0.29$, $\Omega_{\Lambda} = 0.71$  \citep[see][]{WMAP_2013}, and distances are all given in proper kpc.

\section{Data: the MEGAFLOW survey}
\label{sec:data}

The MEGAFLOW sample is based on the combination of Integral Field Unit (IFU) data from the Very Large Telescope (VLT) Multi Unit Spectrograph Explorer \citep[MUSE,][]{Bacon_2010} and
Ultraviolet and Visible Echelle Spectrograph \citep[UVES,][]{Dekker_2000} observations of 22 quasar fields. 
MUSE provides spectral cubes from which galaxies can be detected either from their continuum or 
from their emission lines  (even without continuum) over a large field of view $1 \times 1$ arcmin$^2$. Depending {on} the field, the exposure time spans from 1.7~hr (shallow field) to 11.2~hr (deep field) with a mean value of 3.9~hr. The PSF at 7000~\AA\ is typically $\approx 0.7$~arcsec.
As discussed in \citet{Bouche_2024}, the galaxy sample is 50\% complete down to  $i\approx 25.2$~m$_{\rm AB}$ ($26.0$~m$_{\rm AB}$) for continuum detected galaxies in the shallow (deep) fields, and $\approx 3.7 \times 10^{-18}$~erg~s$^{-1}$~cm$^{-2}$ ($7 \times 10^{-18}$~erg~s$^{-1}$~cm$^{-2}$) for galaxies detected from their [\OII] emission.

While the survey is designed around 79 strong \MgII{}
absorption systems (rest-frame equivalent width \REW > 0.5 \AA) selected from the SDSS catalog of 
\citet{Zhu_Menard_2013}, additional \MgII\ absorptions have been identified in high-resolution UVES spectra down to $\approx$0.05~\AA (3$\sigma$). The final \MgII{} catalog is  composed of 127 \MgII\ absorbers identified in the UVES spectra. Most of these (120) have a redshift $0.3< z < 1.5$ for which the [\OII] doublet falls in the MUSE range.

 In this work, we use the galaxy catalog of the MEGAFLOW catalog (version v2.0beta) from \citet{Bouche_2024} which contains a total of  1998 galaxies with redshifts. Among these galaxies 730 have been identified solely from their emission lines (no continuum) and 1208 are located in the foreground of the 22 quasars.  For more details about the MEGAFLOW survey catalog see \citet{Bouche_2024}. 

\section{Galaxy sample}
\label{sec:isolation_criteria}

In the present work, we quantify the \MgII\ profile around galaxies and measure its dependence on several galaxy properties such as SFR, $z$, and orientation. For that, 
we want to be sure that our absorption/galaxy pairing is unambiguous and therefore focus on isolated galaxies  with a redshift difference relative to the QSO corresponding to a velocity difference $\Delta v_{\mathrm{QSO}} \geq 3000 ~\kms$.
Isolated galaxies can be unambiguously associated with absorption (if present) and are unlikely to have experienced a major interaction in their recent history. We describe below the selection of our sample and the criteria we use to define "isolated" galaxies.

\begin{figure}
	\includegraphics[width=8.5cm]{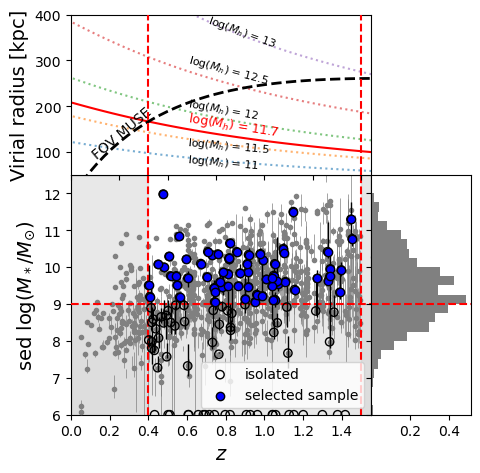}
    \caption{\textbf{Top}: Evolution of the virial radius with redshift for different halo masses compared to the evolution of the size of the MUSE field of view. The full red line represents the virial radius for the median halo mass of our sample (that we choose as our isolation radius). The vertical dashed lines indicate our redshift selection ($0.4 < z< 1.5$) for which the isolation radius is contained within the MUSE field of view.
    \textbf{Bottom}: distribution of the stellar masses for MEGAFLOW galaxies at $z< 1.5$. We consider the MEGAFLOW survey reasonably complete for $0.4 < z< 1.5$ down to $\log(M_\star/\msun) = 9$ (represented by the horizontal dashed line). Our volume limited sample of isolated galaxies is represented by blue dots}
    \label{fig:Fov_completeness}
\end{figure}

First, we select galaxies in $0.4 < z < 1.5$. In this redshift range \MgII\ absorptions can be detected with UVES and galaxies can be detected from [\OII] with MUSE. We only keep galaxies with a redshift confidence ZCONF$>2$ \citep[see the survey paper][for details about the ZCONF score]{Bouche_2024}. Among this redshift range MEGAFLOW is complete in stellar mass down to $\log(M_\star/\msun)= 9.25$ and is 60\% complete in the bin $9 < \log(M_\star/\msun)< 9.25$. We then select galaxies with $\log(M_\star/\msun) > 9$ (for details about the stellar mass estimation see section \ref{sec:galaxy_properties}).
Second, in order to select isolated galaxies we define a maximum projected distance $r_{\rm max}$ around the quasar LOS in which we require the galaxies to be alone in a $\pm 500 ~\kms$ redshift range. Instead of using an arbitrary distance such as 100~kpc or 150~kpc for $r_{\rm max}$, here we adopt a more physical approach that takes into account the evolution of halo size with redshift, namely we choose $r_{\rm max}$ to be the virial radius of a dark matter halo. Figure \ref{fig:Fov_completeness} (top panel) shows the redshift evolution of the virial radius for different halo masses. For the median stellar mass ($\log(M_\star/\msun) = 9.8$) of our sample, the corresponding  halo mass is $\log(M_{\rm h})= 11.7$ using   the $z\approx1$ stellar-to-halo mass relation from \citet{Behroozi_2019}, such that $r_{\rm max}$ is  105~kpc at $z = 1.5$ and $\approx$ 165~kpc at $z = 0.4$ shown  as the red line in Fig.~\ref{fig:Fov_completeness}(top). This figure also shows that $r_{\rm max}$ is inside the FOV at all redshifts. This isolation distance defines a cylinder, never cropped by the FOV and centered on the LOS in which we require that there is a single galaxy. However, this does not mean that there is no neighbour around that galaxy outside the isolation distance (a neighbour could be present outside of the FOV for instance). In addition, as shown in \citet{cherrey_2023}, \MgII\ absorption around groups could extend much further than around field galaxies. For that reason, we do not consider cases when there are 5 or more galaxies in the field of view (N$_{\rm FOV} \geq 5$) within a $\pm 500 ~\kms$ redshift range, (even if one of them is isolated as defined above). Finally, as we aim at studying \MgII\ absorption, we remove cases of galaxies having a redshift at which {the MgII doublet is truncated or not covered by UVES}.

Hereafter, we define our isolated galaxy sample to be the volume-limited sample of isolated galaxies with $\log(M_\star/\msun) > 9$ at $0.4<z<1.5$. We summarize below the selection criteria that we apply successively on the MEGAFLOW catalog to form this sample:

\begin{itemize}
    \item $0.4 < z < 1.5$ (1124 gals)
    \item and ZCONF $\geq$ 2 (960 gals)
    \item and $z_{\rm gal}$ $<$ $z_{\rm QSO}$ with $\Delta v_{\mathrm{QSO}}\geq3000$ \kms (905 gals) 
    \item and $N_{\rm FOV}$ < 5 to avoid groups
    \item and alone within $R_{\rm vir}(M_h,z)$ (144 galaxies)
    \item and good UVES coverage (128 galaxies)
    \item and $\log(M_\star/\msun) > 9$ (66 galaxies = \textit{isolated galaxy sample})
\end{itemize}

With these criteria, our sample of isolated galaxies is made of 66 isolated galaxies, with a median mass of $\log(M_\star/\msun) = 9.8$ and a median redshift of $z = 0.91$.

\section{Galaxy properties}
\label{sec:galaxy_properties}
\subsection{Stellar masses}
For the foreground galaxies having a magnitude $r < 26$~mag, we estimate the stellar masses using the Spectral Energy Distribution (SED) fitting code \texttt{CONIECTO} \citep[For details see][]{Zabl_2016,Bouche_2024}. It combines stellar continuum emission from the BC03 model \citep[][]{Bruzual_Charlot_2003} with nebular emission using the method of \citet{Schaerer_Barros_2009}, under the assumption of a \citet{Chabrier_2003} IMF and a \citet{Calzetti_2000_Dust} extinction law.
The bottom panel of Figure \ref{fig:Fov_completeness} shows the stellar mass as a function of redshift for the foreground galaxies at z < 1.5 in the MEGAFLOW data. The typical statistical error on $\log(M_\star/\msun)$ is $\approx 0.15$~dex.  {The stellar mass is not computed for galaxies with a $r$ magnitude $>26.5$.}

\subsection{Star Formation Rate}
To estimate the Star Formation Rate (SFR) of the galaxies We use the empirical relation from \citet{gilbank_2011}. This relation has been derived by calibrating, as a function of mass, the [\OII] luminosity against \Ha\ based SFR corrected from dust extinction using an average Balmer decrement:

\begin{equation}
    SFR = \frac{L([\OII]_{\rm obs})/3.8\times10^{40}\rm erg/s}{a \tanh[(x-b)/c]+d},
\end{equation}
with $a = -1.424$, $b = 9.827$, $c = 0.572$, $d = 1.70$ and $x = \log(M_\star/M_\odot)$. The [\OII] detection limit imposes a SFR limit of $\approx 0.01~\msun/$yr at $z = 1$. The median SFR of our sample is $3.1~\msun/$yr.

\subsection{\galpak\ modeling results}
In order to derive the galaxies' morphological and kinematic properties we use the \galpak\ algorithm \citep{Bouche_2015}. \galpak{} is a Bayesian code that fits a parametric disk model directly to the observed 3D data ($x$, $y$, $\lambda$). The disk model assumes an analytic form for the rotation curve $v(r)$, in our case we use the \textit{Universal Rotation Curve} (URC) from \citet{Salucci_1997}. For that reason we denote hereafter the \galpak{} fits as URC fits.  
\galpak{} then convolves the disk model with the instrument Point Spread Function (PSF) and Line Spread Function (LSF) such that it yields the intrinsic parameters of the galaxy. The derived parameters of the disk model are position, inclination, azimuthal angle, effective radius, maximum circular velocity, thermal velocity dispersion, total flux, Sérsic index, [\OII] line ratio, turnover radius of the rotation curve and shape parameter of the rotation curve. 
The advantage of a 3D fit is that it takes into account all the information present in the data, even if the signal-to-noise ratio (S/N) is low for each individual spaxel.

 We first run \galpak\ on all the 66 isolated galaxies of our sample (as defined in section \ref{sec:isolation_criteria}). For most of them (63) we use the [\OII] doublet as it is the brightest emission line. We use the [\OIII] $\lambda$5007 and H$\beta$ emission lines for respectively two and one galaxies, for which [\OII] is fainter. In this work, we focus on isolated galaxies for which the emission S/N is sufficient to obtain a robust morphology and kinematic model with \galpak. For that reason, we removed galaxies with S/N$< 4$. After detailed case-by-case visual inspection of the \galpak fits and comparison of modeled velocity maps versus observed velocity maps produced with CAMEL \citep{Epinat_2009_camel1, Epinat_2012_camel2}, we removed 2 additional galaxies that appear to be irregular and hence cannot be modeled by a disk. Finally, we find that the morphology and kinematics have been convincingly determined for 60 out of our 66 isolated galaxies.

\begin{figure}
\includegraphics[width=7.5cm]{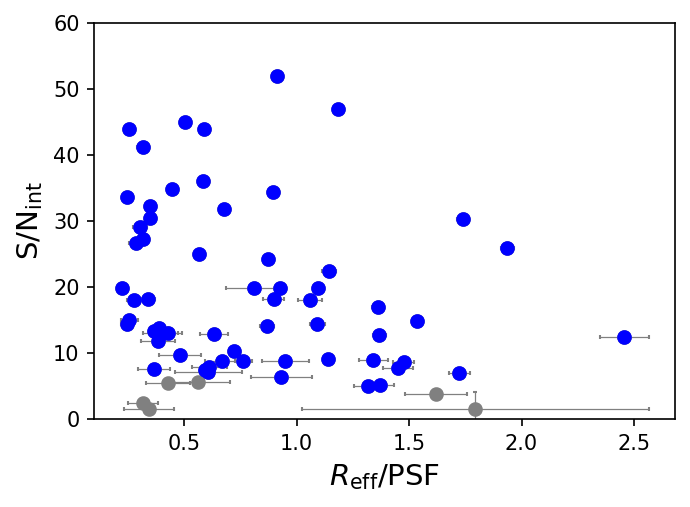} 
    \caption{Integrated S/N from \texttt{pyplatefit} versus effective radius divided by PSF FWHM for the 66 isolated galaxies of our isolated galaxy sample. Galaxies for which \galpak fits\ seem reliable are represented by blue dots. Galaxies with bad fits are represented by grey dots. Galaxies with bad \galpak fits tend to have a low integrated S/N.}
    \label{fig:RePSF_SNR}
\end{figure}

Among the 60 galaxies for which the morphological and kinematic parameters have been determined with \galpak, 5 galaxies are dispersion dominated as their circular velocity to velocity dispersion ratio at twice the effective radius, $v/\sigma(2R_{1/2})$, is below 1. Finally, out of the 55 rotation-dominated galaxies, 46 have an inclination angle $i > 30$ which is sufficient to accurately estimate the azimuthal angle $\alpha$ between the major axis and the LOS (see Figure \ref{fig:sketch_angle}). Note that
the distribution of inclinations for the 55 isolated rotation-dominated galaxies follows the expected $\sin i$ relation
 (a KS-test gives a p-value of 0.25) and that the distribution of azimuthal angles $\alpha$ for the 46 galaxies with a sufficient inclination is consistent with a flat distribution (a KS-test gives a p-value of 0.56). These distributions are shown in  Fig.~\ref{fig:inclination_distrib_isolated}.

\section{Properties of gaseous halos}
\label{sec:MgII_halos}

Now that we have selected a sample of isolated galaxies, we study their cool gas profile by looking at \MgII\ absorption in the quasar sight-lines. 
In particular, we quantify the \MgII\ rest-frame equivalent width and the covering fraction as a function of the impact parameter in sections \ref{sec:W-b} and \ref{sec:fc} respectively. We then investigate how the \MgII{}  profile depends on redshift and SFR in section \ref{sec:z,M,sfr} and on orientation and inclination in section \ref{sec:alpha,i}. 

\subsection{The rest-frame equivalent width -- impact parameter relation (\REW - $b$)}
\label{sec:W-b}

We want to establish how the \MgII\ absorption strength varies as a function of the impact parameter. Hereafter, we consider that a galaxy is associated with a \MgII\ absorption system if the velocity difference between the two is $\leq 500 ~\kms$ even though 90\%\ of the velocity differences are $\leq$ 200 (100)~\kms\ at impact parameters $b<100$ (50)~kpc as discussed in the survey paper \citep{Bouche_2024}. As presented in the same paper, the $3\sigma$ detection limit for \MgII\ absorption does not evolve significantly between redshift 0.4 and 1.5 with a $75^{\rm th}$ percentile value of $\approx 0.075$~\AA\ over our 22 fields. We consider this value as an upper limit of detection when no \MgII\ absorption is detected.

The left panel of Fig.~\ref{fig:fc_det_lim} shows \REW{} as a function of the impact parameter $b$ for our isolated galaxy sample.
Below 50 kpc, all $\log(M_\star/\msun) > 9$ isolated galaxies (21) are associated with \MgII\ absorption having \REW\ $>0.1$\AA. Between 50 and 100 kpc, there is a mix of galaxies with and without absorption above this \REW\ value. Above 100 kpc, the vast majority of galaxies are not associated with absorption above our detection limit and the few cases of detected \MgII\ absorption have \REW\ $< 0.2$ \AA.

To model the relation between absorption equivalent width and impact parameter, we perform a fit similar to that proposed by \citet{Chen_2010} and  \citet{Dutta_2020}, assuming a log-linear relation of the form:
\begin{equation}
\log{W^{2796}_r}(b) = a_0 + a_1 \times b,
\label{eq:W-b_simple}
\end{equation}
where $b$ is the impact parameter.  We fit by maximizing the following likelihood:
\begin{equation}
    \begin{split}
        \mathcal{L}(W) & =  \left[ \prod_{i=1}^n \frac{1}{\sqrt{2 \pi}\sigma_i} \exp \left(-\frac{1}{2} \left[\frac{W_i - W(b_i)}{\sigma_i} \right]^2\right)\right]  \\
        & \times \left[ \prod_{i=1}^m \int_{-\infty}^{W_{i}}  \frac{dW'}{\sqrt{2 \pi}\sigma_i} \exp \left(-\frac{1}{2} \left[\frac{W' - W(b_i)}{\sigma_i} \right]^2\right)\right],
    \end{split}
    \label{eq:split}
\end{equation}
where $W_i \equiv \log{W^{2796}_{r,i}}$ and $\sigma_i$ its uncertainty associated. The first factor in Eq.~\ref{eq:split} corresponds to the likelihood of the $n$ points that have detected \MgII\ absorption and the second factor corresponds to the likelihood of the $m$ points that do not have \MgII\ absorption detected but only an upper limit on $W_{i}$.

The uncertainty $\sigma_i$ to $W_i$ is  composed of our measurement uncertainty $\sigma_{m_i}$ and an intrinsic  scatter $\sigma_{c}$ added in quadrature:
\begin{equation}
    \sigma_i^2 = \left(\frac{\sigma_{m_i}}{\ln(10) W^{2796}_{r,i}}\right)^2 + \sigma_{c}^2.
\end{equation}
The intrinsic scatter $\sigma_{c}$ is added as a free parameter and fitted simultaneously. We perform this fit for our isolated galaxy sample. The result is presented in Table \ref{tab:model_params} and shown in Figure \ref{fig:fc_det_lim} along with similar fits from \cite{Nielsen_2013, Dutta_2021, Huang_2021}.  Our result is mostly consistent with the fit from \citet{Nielsen_2013} obtained for a sample of 182 isolated galaxies at $0.07<z<1.12$. They define an isolated galaxy as a galaxy that has no neighbour within 100 projected kpc and $\pm 500$ \kms, which is compatible with our definition in many cases. They also probe a similar impact parameter range. The result from \cite{Dutta_2021} using the MAGG \citep{Lofthouse_2020, Dutta_2020, Fossati_2021} and QSAGE \citep{Bielby_2019, Stott_2020} samples differs significantly from what we obtain. This could be due to the fact that their galaxies are not isolated (when multiple galaxies are present they consider the one closest to the line of sight), that could explain their higher equivalent width at high $b$ values. Their sample is also probing a different impact parameter range, with very few cases at small $b$ (below 50 kpc), which could explain the discrepancy at these distances. They do not apply any selection on stellar mass, so their sample have a lower median mass ($\log(M_\star/\msun) = 9.3$) than our ($\log(M_\star/\msun) =9.8$). The result from \citet{Huang_2021}, based on 211 isolated galaxies from SDSS, is broadly consistent with our result even though the fitting function do not have the same form, their redshift range is lower ($z<0.5$) and their isolation criterion is different with no neighbour within 500 projected kpc and $\pm 1000$ \kms.

\subsection{Covering fraction}
\label{sec:fc}

We now turn to compute the \MgII{} covering fraction $f_{\rm c}$. The covering fraction is defined as the probability $P$ to detect a \MgII\ absorption above a \REW\ threshold, at a given projected distance from a galaxy and within a redshift range $\Delta v$. Given that there are only two possible outcomes (above some threshold there is an absorption ($Y=1$) or not ($Y=0$)), several methods exist to estimate the covering fraction
\citep[e.g.][]{Wilde_2021,Schroetter_2021,Huang_2021}. The most popular method relies on counting the outcomes $Y$ in bins of $b$ \citep[e.g.][]{Nielsen_2013_Magiicat}. Here, we compute the differential covering fraction following the method described in \citet{Schroetter_2021} and \citet{cherrey_2023}, which has the advantage that it does not require any binning. Briefly, it consists in a Bayesian logistic regression where the covering fraction follows  the logistic form:

\begin{equation}
    P(Y = 1) \equiv L(t) =  \frac{1}{1+\exp{(-t)}},
	\label{eq:logistic}
\end{equation}
where $P$ is the probability of detection $Y=1$ and $t$ is a function of the independent variables $X_i$ and of the model parameters $\theta$. We assume, as a first step, that the covering fraction is governed by the impact parameter $b$ to the first order and follows the form:

\begin{equation}
    t = f(X_i, \theta) = k_1 (\log{b} - k_0).
	\label{eq:t_simple}
\end{equation} 
where $k_0$ is the zero-point corresponding to the log distance at which $P=0.5$, and $k_1$ is the steepness of the transition of $P$ from 0 to 1.
We fit the parameters $k_0$, and $k_1$ using a Monte Carlo Markov Chain (MCMC) procedure along with a Bernouilli likelihood optimized on 9000 steps. We perform the optimization with the python module \texttt{pymc3}  \citep{Hoffman_2011, Salvatier_2015}. Hereafter, we denote $b_{50}$ as the impact parameter corresponding to a 50\%\ probability of having an absorption, i.e. $f_{\rm c}=50\%$.

\begin{figure*}
    \includegraphics[width=8.5cm]{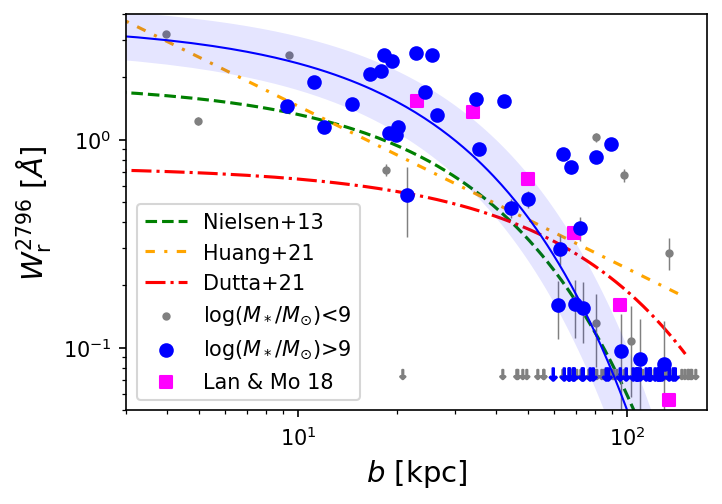} 
	\includegraphics[width=8.5cm]{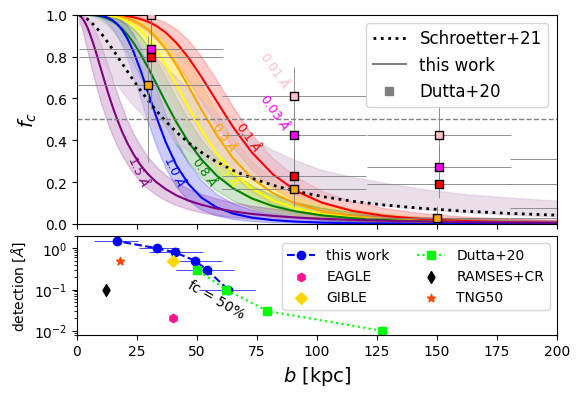} 
    \caption{
    \textbf{Left:} \REW-$b$ relation for the sample of isolated galaxies. The blue dots represent our isolated galaxy sample, small grey dots represent $\log(M_\star/\msun)<9$ isolated galaxies. The blue line is the \REW-$b$ fit for $\log(M_\star/\msun)>9$ isolated galaxies and the shaded area shows the 1-$\sigma$ interval. \REW\-$b$ fits from \citet{Dutta_2020, Nielsen_2013, Huang_2021} are also represented along with the binned values from \citet{Lan_2018}.
    \textbf{Right:} \MgII{} covering fraction as a function of impact parameter for the 66 $\log(M_\star/\msun) > 9$ isolated galaxies for different $W_r$ thresholds. Shaded areas are 1-$\sigma$ uncertainties. The dotted line represents the covering fraction computed by \citet{Schroetter_2021} for MEGAFLOW for a $0.1$\AA\ detection limit for primary galaxies (no mass selection). Squares represent the differential covering fraction from \citet{Dutta_2020} for different $W_r$ thresholds.
    The lower panel shows the impact parameter $b$ corresponding to the 50\% covering fraction as a function of the $W_r$ threshold value for our sample (dashed blue line) and interpolated from the covering fraction values from \citet{Dutta_2020} (dotted green line). We also show the $b_{50}$ from \citet{DeFelippis_2021} (TNG50), \citet{DeFelippis_2024} (RAMSES with cosmic rays), \citet{Ho_2020} (EAGLE) and \citet{Ramesh_Nelson_2024} (GIBLE at z=1).
    }
    \label{fig:fc_det_lim}
\end{figure*}

We apply this method (Eqs.~\ref{eq:logistic}--\ref{eq:t_simple}) to compute the \MgII{} covering fraction as a function of impact parameter, $f_{\rm c}(b)$, for our sample
and investigate its dependence with respect to the \REW{} threshold $W_{\rm t}$ by computing $f_{\rm c}$ for different values of $W_{\rm t}$: 0.1~\AA, 0.3~\AA, 0.5~\AA, 0.8~\AA, 1.0~\AA, and 1.5~\AA{} (see Table \ref{tab:fc_params}). We show in Figure \ref{fig:fc_det_lim} (upper right panel)   $f_{\rm c}(b)$  for these different threshold values. We observe a significant shift of $k_0$ with respect to $W_{\rm t}$, with almost no slope variation $k_1$. We also observe for our sample a steeper slope than \citet{Schroetter_2021} (dotted line) and \citet{Dutta_2020} (solid squares), probably due to our stricter pre-seletion of isolated galaxies and our stellar mass selection (see discussion in section \ref{sec:discussion} below).

The lower right panels shows that $b_{50}$ decreases almost linearly with the threshold value, varying from $\approx 65$~kpc at $\REW>0.1$~\AA\ to $\approx 20$~kpc at $\REW>1.5$~\AA. These measurements are consistent with and complementary (in impact parameter coverage) to the results from \citet{Dutta_2020}. We also show that $b_{50}$ values computed from TNG50 \citep{DeFelippis_2020} and EAGLE \citep{Ho_2020} simulations are significantly lower than our measurements for similar threshold values. We note that the $0.1$\AA\ covering fraction derived by \citet{Schroetter_2021} for primary galaxies in MEGAFLOW is lower than the covering fraction for our  sample. This is mainly due to the mass selection that we apply to obtain a volume limited sample. If we do not apply this criterion, we obtain a similar covering fraction (see appendix \ref{sec:mass_dependence_appendix} for details).

One can argue that the pre-selection of the quasar fields in MEGAFLOW based on the presence of $>3$ strong absorption systems (\REW\ $>0.5$~\AA) could introduce a bias in the results, especially in the covering fraction. \citet{cherrey_2023} presented a method to estimate the magnitude of this bias, which we outline here. The idea is to simulate several quasar fields, randomly populated by galaxies. Each galaxy could be associated or not with an absorption ($\REW>0.1$~\AA) in the quasar spectrum, that in some cases could be strong ($\REW>0.5$~\AA). To mimic realistically the presence of these absorptions, the covering fractions curves down to $\REW>0.1$~\AA\ and to $\REW>0.5$~\AA\ are needed. Here we use the covering fraction computed above. The fields with $>3$ strong absorption systems are then selected to reproduce the MEGAFLOW selection. Finally, the covering fraction is recomputed on these selected fields and compared with what would be obtained for randomly selected fields in order to estimate the bias caused by field selection. For 50 fields of 500 $\times$ 500~kpc containing $\approx 50$ galaxies each, we obtain 21 pre-selected fields (similar to MEGAFLOW). The $b_{50}$ at 0.1~$\AA$ is 66$\pm 5$~kpc (error is 2$\sigma$) for the selected fields versus $63 \pm 5$~kpc for the random fields. The estimated bias due to field pre-selection would hence be minor compared to the statistical uncertainties.

\subsection{Dependence on redshift, stellar mass and SFR}
\label{sec:z,M,sfr}

\begin{figure*}
	\includegraphics[width=18cm]{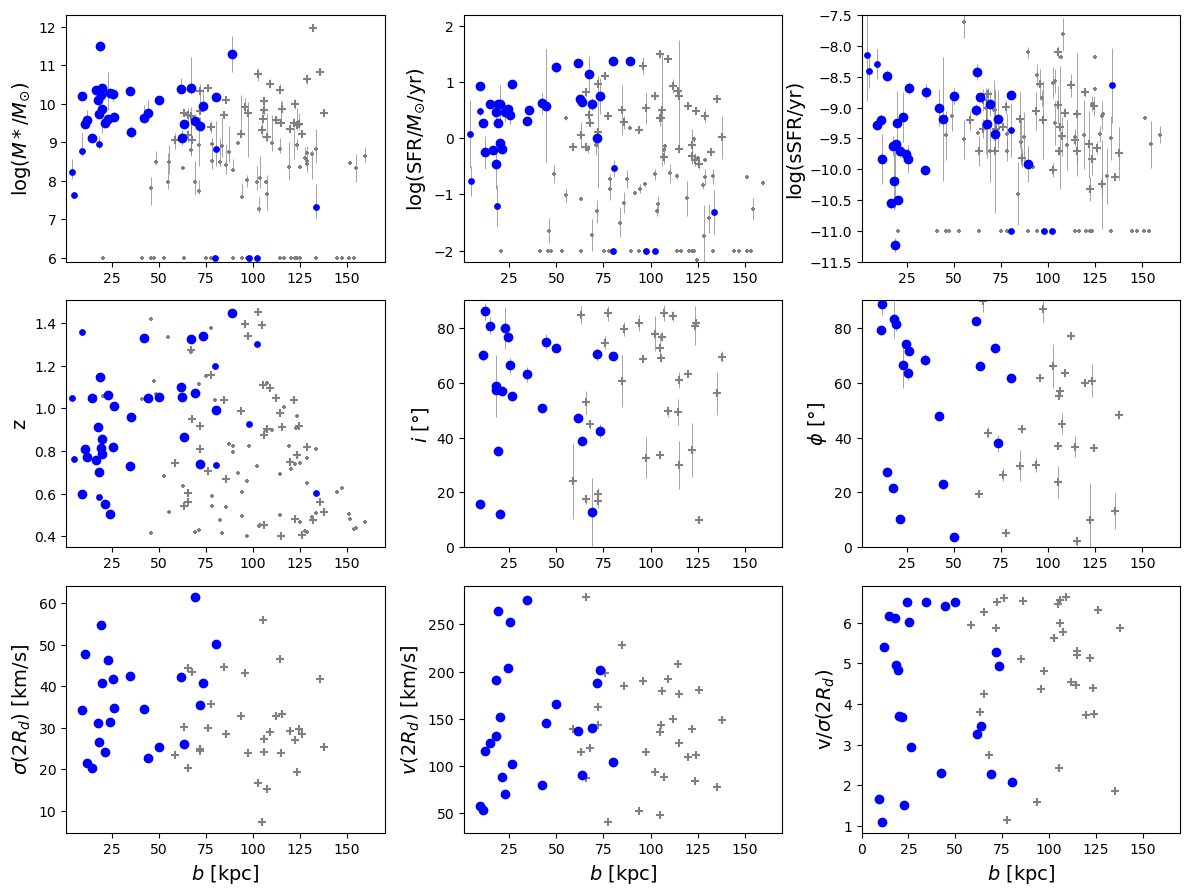} 
    \caption{From left to right and top to bottom: log($M^*$), log(SFR), log(sSFR), redshift, inclination $i$, azimuthal angle $\alpha$, velocity dispersion at $2\;R_d$, circular velocity at $2\;R_d$ and $v/\sigma$ at $2\;R_d$ as a function of the impact parameter for the isolated galaxies. Galaxies associated with absorption {with \REW>0.1 \AA} are represented by blue dots. Galaxies with no absorption detected {or absorptions with \REW<0.1 \AA} are represented by {gray crosses}. For $M_\star$, SFR, sSFR and z, we show all 128 isolated galaxies (the 66 $\log(M_\star/\msun)>9$ isolated galaxies are represented with big markers and the 62 $\log(M_\star/\msun)<9$ isolated galaxies are represented by small ones). For inclination, velocity dispersion, circular velocity and $v/\sigma$ we only show rotation-dominated isolated galaxies (55) and for $\alpha$ those with $i>30$° (46).
    }
    \label{fig:2D_plots}
\end{figure*}

The impact parameter is the most important parameter determining the strength of the absorption and the covering fraction. However, other parameters could also play a role. We investigate the impact of some of them in this section. Figure \ref{fig:2D_plots} shows how the absorption strength depends on a number of properties of the galaxies in our sample. It appears in particular that redshift, star formation rate and stellar mass are correlated with \MgII\ absorption strength and extent.

\begin{figure}
    \includegraphics[width=8.5cm]{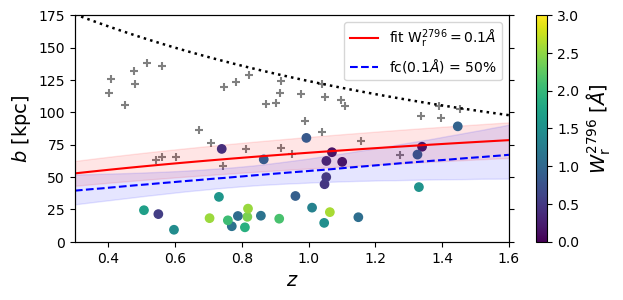} 
	\includegraphics[width=8.5cm]{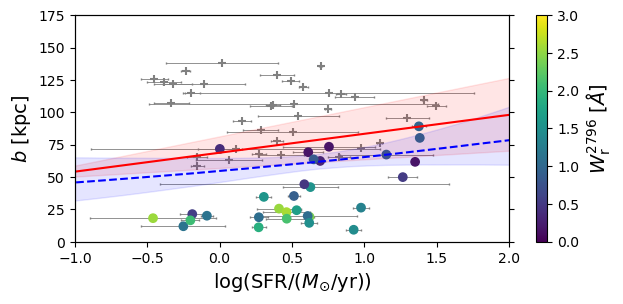} 
    \includegraphics[width=8.5cm]{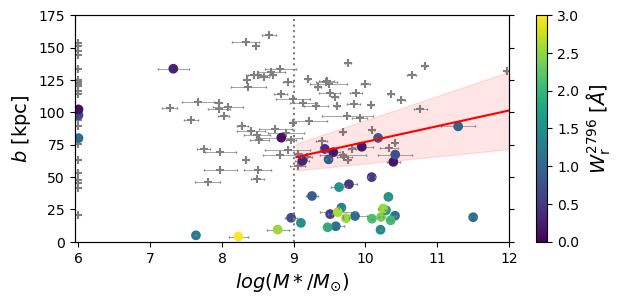} 
    \caption{ {Top and middle panels present respectively the fitted dependence of \MgII\ absorption with $z$ and log(SFR) for our isolated galaxy sample. The bottom panel shows how it converts in terms of $M_\star$ for the isolated galaxy sample extended to low mass galaxies (without the $\log(M_\star/\msun)>9$ selection cut)}.  {On the top and middle panels} the full red lines and the blue dashed lines represent respectively the impact parameter at which the \REW reaches $0.1$\AA\ and the  50\% covering fraction reaches $0.1$\AA\ according to the fits. Shaded areas represent 1-$\sigma$ intervals. We take the other parameters constant with the following values: $z = 1$, log(SFR/($M_\odot~yr^{-1}$)) = 0.  {On the bottom panel the red line is converted from the SFR relation using the main sequence from \citet{Boogaard_2018}.}   {On the three panels the } associated \MgII\ absorption strength is color coded  {for galaxies with \REW>0.1 \AA}.  {Galaxies not associated with absorptions or associated with absorptions with \REW<0.1 \AA} are represented by grey crosses.  {On the top panel} the black dotted line represents the redshift evolution of the virial radius for a halo of mass $10^{11.7}M_h$ (used as our isolation distance).  {On the bottom panel, galaxies with no stellar mass estimation are represented at $\log(M_\star/\msun) = 6$}}
    \label{fig:multi_param_fit}
\end{figure}

In order to investigate these dependencies, we take them into account for the fit of the \REW-$b$ relation and the fit of the covering fraction. SFR and $M_\star$ are tightly correlated through the main sequence, so, in order to limit the uncertainties in the fits we only focus here on the redshift and SFR dependence. In consequence, we replace Eq.~\ref{eq:W-b_simple} by:

\begin{equation}
\begin{split}
\log{W^{2796}_r} = & a_0 + a_1 b/\rm{kpc} + a_2 \log(1+z) + a_3 \log(\rm{SFR}/(\msun/yr))
\end{split}
\label{eq:W-b_multi}
\end{equation}
And for the covering fraction, we replace Eq.~\ref{eq:t_simple} by :

\begin{equation}
\begin{split}
    t = f(X_i, \theta) = & k_1 (\log{(b/\rm{kpc})} - k_2 \log(1+z) \\ & - k_3 \log(\rm{SFR}/(\msun/yr)) - k_0).
\end{split}
	\label{eq:t_multi}
\end{equation} 
We perform these new fits on our isolated galaxy sample.
The fitted parameters for the \REW-$b$ relation and for the covering fraction are presented in Table \ref{tab:model_params} and Table \ref{tab:fc_params} respectively and the resulting fits are shown in Figure \ref{fig:multi_param_fit}.
Despite uncertainties compatible with a flat evolution, we observe a tentative decreasing trend of \REW and $b_{50}$ toward low redshifts. Such result give hints that \MgII\ halos do not follow the growth of DM halos along cosmic time (see discussion below).

{The bottom panel of \ref{fig:multi_param_fit} shows how the SFR dependence of \REW converts into stellar mass dependence. For that we use the main sequence fitted by \citet{Boogaard_2018} and inject it in equation \ref{eq:W-b_multi}.}

\begin{table*}
\centering
\begin{threeparttable}
\caption{\REW -$b$ relation (eq. \ref{eq:W-b_simple}, \ref{eq:W-b_multi}) best-fit  parameters for the isolated galaxy sample, the high alpha and low alpha sub-samples. The fitted equation is: $\log{W^{2796}_r} = a_0 + a_1 b/\rm{kpc} + a_2 \log(1+z) + a_3 \log(\rm{SFR}/(\msun/yr))$ with an intrinsic scatter $\sigma_c$. The uncertainties are 1-$\sigma$}.  
\label{tab:model_params}
\begin{tabular}{c c c c c c} 
 \hline
 & $a_0$ & $a_1$ & $a_2$ & $a_3$ & $\sigma_c$\\ 
 \hline
 full sample & $0.56 \pm 0.13$ & $ -0.02 \pm 0.002$ & & & $0.40 \pm 0.06$\\ 
 \\
  full sample & $-0.04 \pm 0.10$ & $ -0.022 \pm 0.002$ & $1.90 \pm 0.15$ & $0.33 \pm 0.15$ & $0.41 \pm 0.06$\\ 
 \\
  $\alpha > 60$ & $0.65 \pm 0.13 $ & $-0.018 \pm 0.002$ & & & $0.31 \pm 0.06$\\
  \\
  $\alpha < 30$ & $ 0.69 \pm 0.20$ & $-0.027 \pm 0.005$ & & & $0.24 \pm 0.09$\\
 \hline
\end{tabular}
\end{threeparttable}
\end{table*}

\begin{table*}
\centering
\begin{threeparttable}
\caption{Covering fraction parameters (eq. \ref{eq:t_simple}, \ref{eq:t_multi}) for the isolated sample, the high alpha and low alpha sub-samples. The fitted equation is $t = f(X_i, \theta) = k_1 (\log{b/\rm{kpc}} - k_2 \log(1+z) - k_3 \log(\rm{SFR}/(\msun/yr)) -  k_0)$. The uncertainties are 1-$\sigma$.}
\label{tab:fc_params}
\begin{tabular}{c c c c c c} 
 \hline
 & $k_0$ & $k_1$ & $k_2$ & $k_3$ & $b_{50}$ [kpc]\\ 
 \hline
 full sample at 0.1 \AA\ & $1.80 \pm 0.04$ & $ -10.31 \pm 2.64$ & & & 63\\ 
 \\
  full sample at 0.3 \AA\ & $1.74 \pm 0.05$ & $ -9.89 \pm 2.51$ & & & 55\\ 
 \\
  full sample at 0.5 \AA\ & $1.70 \pm 0.06$ & $ -9.39 \pm 2.31$ & & & 50\\ 
 \\
  full sample at 0.8 \AA\ & $1.61 \pm 0.06$ & $ -7.94 \pm 1.81$ & & & 41\\ 
 \\
  full sample at 1.0 \AA\ & $1.52 \pm 0.06$ & $ -10.21 \pm 2.44$ & & & 33\\ 
 \\
  full sample at 1.5 \AA\ & $1.23 \pm 0.13$ & $ -4.64 \pm 1.44$ & & & 17\\ 
 \\
  full sample at 0.1 \AA\ & $1.51 \pm 0.21$ & $ -10.14 \pm 2.37$ & $0.77 \pm 0.74$ & $0.08 \pm 0.09$ & 32\tnote{*}\\ 
 \\
  $\alpha > 60$ at 0.1 \AA\ & $1.87 \pm 0.09 $ & $-7.25 \pm 3.35$ & & & 75 \\
  \\
  $\alpha < 30$ at 0.1 \AA\ & $1.68 \pm 0.14$ & $-7.09 \pm 3.44$ & & & 48\\
 \hline
\end{tabular}
\begin{tablenotes}
       \item [*] given for z = 1, log(SFR/($M_\odot yr^{-1}$)) = 0, log($M_\star/\msun$) = 9.0.
\end{tablenotes}
\end{threeparttable}
\end{table*}

\subsection{Dependence on the orientation and inclination angles}
\label{sec:alpha,i}

\begin{figure}
    \includegraphics[width=8.cm]{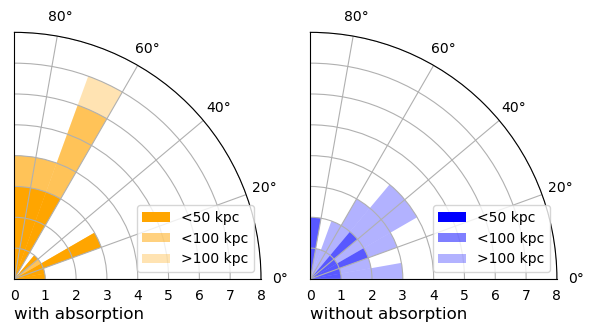} 
    \caption{Number of galaxies as a function of the azimuthal angle for the 46 rotation dominated isolated galaxies with $i>30$° and with an associated absorption (left) or without an associated absorption (right).}
     \label{fig:alpha_histogramms}.
\end{figure}

Several previous observations suggested that the CGM show an azimuthal bimodality \citep[e.g][]{ Bouche_2012, Kacprzak_2012, Schroetter_2019}. Indeed, galactic winds with higher metallicity are preferentially ejected along the minor axis of galaxies \citep[e.g][]{Rubin_2014, Schroetter_2016, Guo_2023, Zabl_2021} while accretion rather happens preferentially in co-rotating disks along the galactic plane \citep[e.g][]{Ho_2017, Zabl_2019, Nateghi_2024a, Das_2024}. For that reason, one might expect \MgII\ absorption to depend on both inclination, $i$, and azimuthal angle, $\alpha$. In Figure \ref{fig:2D_plots}, we see that galaxies with low inclination values have weaker absorption than galaxies with high inclination. Indeed there is only one case of absorption between 50 and 100~kpc for $i<30$°. In Figure \ref{fig:2D_plots}, we also see that \MgII\ absorptions are detected out to larger distances for high $\alpha$ than for low $\alpha$.
In order to better visualize the $\alpha$ dependence,
Figure \ref{fig:alpha_histogramms} presents the distribution of azimuthal angle for galaxies with and without \MgII{} absorption systems. We see that the absorption cases are preferentially located along the minor axis while non absorption cases are preferentially observed for intermediate azimuthal angles ($30 < \alpha <60$).

\begin{figure*}
    \includegraphics[width=8.5cm]{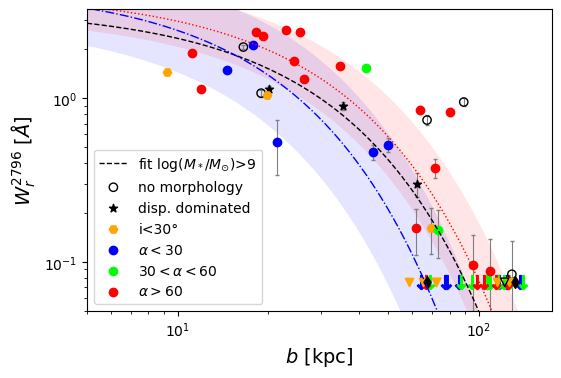} 
    \includegraphics[width=8.5cm]{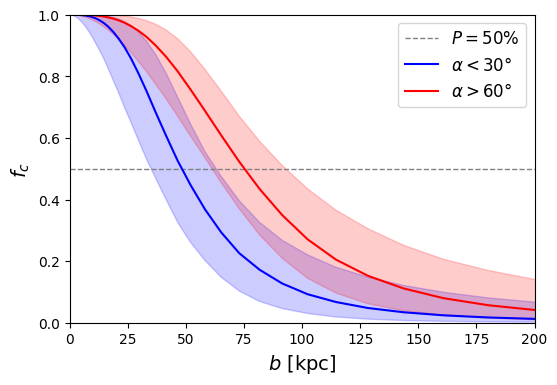} 
    \caption{\textbf{Left:} \REW\-$b$ relation for the 66 isolated galaxies from our sample. The best log linear fits of are shown for the whole sample as well as for the low-$\alpha$ and high-$\alpha$ sub-samples. The mid-$\alpha$ sub-sample is not sufficiently populated at small $b$. Shaded areas are 1-$\sigma$ intervals. \textbf{Right:} The $0.1$\AA\ \MgII\ covering fraction for the low-$\alpha$ and high-$\alpha$ sub-samples, represented by the blue and red lines, respectively. Shaded areas are 1-$\sigma$ intervals.}
     \label{fig:W_b_isolated_alpha_split}
\end{figure*}

To quantify these dependences, we split the rotation-dominated galaxies from our sample (55 in total) into four sub-samples: a low inclination sub-sample ($<30$°, 9 galaxies) for which the azimuthal angle could not be determined robustly, and three sub-samples with $i >30$°: low $\alpha$ ($<30$°, 13 galaxies), intermediate $\alpha$ ($30<\alpha<60$°, 13 galaxies) and high $\alpha$ ($>60$°, 20 galaxies). The \REW\ - $b$ relation for these sub-samples is shown in the left panel of Figure \ref{fig:W_b_isolated_alpha_split}.
It appears that the high $\alpha$ sample, for which the line of sight roughly intersects the galaxy's minor axis, is associated with higher \REW values than the low $\alpha$ sample, for which the line of sight roughly intersects the galaxy's major axis. The only two cases of absorption at impact parameter above 80 kpc are for high-$\alpha$ galaxies. Low-$\alpha$ galaxies seems to have slightly weaker absorption than high-$\alpha$ galaxies. This is particularly visible in the range 50 - 100 kpc where there is only one case of absorption for low $\alpha$ versus 5 for high $\alpha$. For the intermediate-$\alpha$ sample, it is difficult to draw conclusions as there are very few cases at small impact parameter. Nonetheless in the range 50 - 100 kpc, there is only one case of absorption associated with a galaxy with intermediate $\alpha$.

Finally, we compute the covering fraction using equation \ref{eq:t_simple} for the low-$\alpha$ and high-$\alpha$ samples (not for the intermediate-$\alpha$ and low-inclination samples as they are not sufficiently populated at low $b$). The best fitting parameters are presented in Table \ref{tab:fc_params} and the covering fractions are shown in the right panel of Figure \ref{fig:W_b_isolated_alpha_split}. The high-$\alpha$ covering fraction appears to be higher than the low $\alpha$ covering fraction, suggesting that \MgII\ is more abundant in outflows than in accretion disks.  

\section{Discussion}
\label{sec:discussion}

We have shown here that \MgII\ absorption is ubiquitous for a sample of isolated star forming galaxies, with $\log(M_\star/\msun)> 9$ and $0.4 < z < 1.5$. Indeed, all galaxies in our sample within an impact parameter of 55 kpc have associated \MgII\ absorption.
\citet{Rubin_2014} and \citet{Schroetter_2019} suggested that out-flowing gas may not be able to escape galactic potential well for sufficiently massive galaxies and would then be re-accreted along the galactic plane in a "fountain" mechanism \citep{Fraternali_2008} after reaching distances of $\approx $ 50 kpc. Our results are consistent with this scenario. First, because \MgII\ absorption clearly drops between $\approx$ 50 and 100 kpc (with only two cases of absorption beyond 100 kpc for our main isolated galaxy sample). Second, because we surprisingly find several cases of absorptions at large impact parameters (above 80 kpc and up to 150 kpc) for low-mass isolated galaxies ($\log(M_\star/\msun)< 9$), even though we observe much fewer absorptions in general for this low-mass sample. This could indicate escaping outflows for these lower potential galaxies. It could also just be a misassignment of the absorption due to the presence of another undetected nearby low mass galaxy (as we are not complete below $10^9 \msun$).

{The bottom panel of Figure \ref{fig:multi_param_fit} shows how the fitted \REW\ dependence to SFR converts in $M_\star$ dependence.} In Appendix \ref{sec:mass_dependence_appendix} we also show that the covering fraction is significantly higher for $\log(M_\star/\msun)> 9$ isolated galaxies than for $\log(M_\star/\msun)< 9$ isolated galaxies. This correlation with stellar mass was already observed in previous works in absorption \citep{Huang_2021, Lan_2020, Chen_2010_mgii_extent} or in emission using stacking \citep{Guo_2023, Dutta_2023}. This is also consistent with the positive correlation between \MgII\ absorption and luminosity observed for instance by \citet{Chen_2010, Nielsen_2013}.

 Our results for isolated galaxies complement the results from \citet{Dutta_2020, Dutta_2021} and \citet{cherrey_2023} for groups. They showed that \MgII\ absorption is more probable and more extended for over-dense regions containing several galaxies and occupying more massive DM halos. This effect could be partly explained by the "superposed" \citep{Bordoloi_2011, Nielsen_2018} contribution of the CGM of the different galaxies and by various interactions producing an "intragroup" medium. Here we show that, even for isolated galaxies, more massive halos are more likely to be associated with \MgII\ absorption at a given impact parameter. The anti-correlation between \REW\ and halo mass observed using cross-correlation techniques \citep{Bouche_2006, Lundgren_2009, Gauthier_2009} would then come from the low abundance of massive DM halos compared to light ones. 

 Recently, \citet{Guha_2022} observed that ultra strong absorbers are associated with massive galaxies with intermediate SFR, but not with starburst galaxies as proposed in other works. Our Figure \ref{fig:multi_param_fit} reveals that, at small $b$, galaxies with high SFR (log(SFR/$\msun$/yr)$>1$) do not show particularly high \REW. Nevertheless, it appears that \MgII\ absorption extends further out for high SFR galaxies (up to 100 kpc). This could indicate either that star formation activity is not so bursty or alternatively that the CGM has a relatively slow response to the star formation activity. Indeed, an outflow ejected at a typical velocity of $\approx 150 \kms$~takes $\approx 650$~Myr to reach 100~kpc. 


 Figure \ref{fig:fc_det_lim} (lower right panel) shows the difficulty simulations have in matching the observed \MgII\ covering fraction. 
 \citet{Liang_2016} and \citet{DeFelippis_2024}  {with RAMSES \citep{Teyssier_2002_ramses}}, or \citet{DeFelippis_2021} {with TNG50 \citep{Marinacci_2018_TNG, Naiman_2018_TNG}} show a cold phase that presents large structures with sharp limits. At the opposite, detailed hydro-dynamical models such as \citet{McCourt_2018} or analytical models such as \citet{Faerman_2023} tend to favor a “foggy” cold phase with a high number of small clouds ($<1$~kpc) that cannot be reproduced by current  {cosmological} simulations. As a consequence the improvement of the resolution seems to enhance the presence of cold gas in the CGM \citep{Hummels_2019, VdV_2019}. In Figure \ref{fig:fc_det_lim} bottom right panel we show that the GIBLE simulations, which are high resolution,  {zoom-in}  simulations of TNG50 galaxies  {(with 512 better mass resolution)}, present covering fraction that are closer to our results. The unresolved nature of the CGM could explain the dominance of the hot phase ($T\approx 10^{5-6}$~K) outside of the central region ($\lesssim 20$~kpc) as presented in the temperature maps from \citet{DeFelippis_2024, Ho_2020, Liang_2016}. The implementation of cosmic rays seems to contribute to decrease the temperature \citep{Farcy_2022, Liang_2016, DeFelippis_2024} hence to favor the presence of cool clouds.

 In Figure \ref{fig:2D_plots}, we do not see any clear trend between \MgII\ absorption and sSFR, indicating that the extent and probability of \MgII\ absorption is not strongly impacted by the position of the galaxy relative to the main sequence. In particular, we do not observe that passive galaxies (with low sSFR) are associated with weaker \MgII\ absorption. The positive correlations between \REW\ and stellar mass, and between \REW\ and SFR would then be mainly explained by mass and SFR being correlated along the main sequence.

We also confirmed here the previously reported azimuthal bimodality of \REW\ and we quantified the \MgII\ covering fraction along the minor and major axes for galaxies with a sufficient inclination. We find that the covering fraction along the minor axis is significantly higher (at the $2\sigma$ level) than the covering fraction along the major axis, which is in agreement with the findings from \citet{Kacprzak_2012}. With a large sample of star-forming galaxies at $z\approx 1$ \citet{Lan_2018} also reported that metal absorption is twice stronger along the minor axis within 100~kpc. \citet{Huang_2021} reported a weak enhancement of \MgII\ absorption along the minor axis ($\lesssim 1 \sigma$). This weaker trend could be due to the fact that their sample is composed of galaxies having a lower redshift ($z<0.5$) so at a later stage of their evolution. Despite a low number of cases, we also find that low inclination galaxies have lower \REW, which is consistent with the conclusions from \citet{Nielsen_2015} and with the accretion-outflow model of the CGM.

Our results on the redshift evolution (Fig.~5) are consistent with the positive correlation between \MgII\ absorption and redshift shown in other absorption studies \citep{Nielsen_2013, Lan_2020, Lundgren_2021, Schroetter_2021, Dutta_2021} and emission studies \citep{Dutta_2023}. This result indicates that the CGM (at a given mass) is becoming smaller with cosmic time, which is counter-intuitive as dark matter halos grow with time. It goes against the common assumption that \MgII\ halos evolve with DM halos \citep{Chen_2010, Churchill_2013, Faerman_2023, Huang_2021} and extend to a fixed fraction of the DM halo radius. A possible explanation could come from the relation between SFR and redshift. Indeed, if the extent of \MgII\ halos is driven by the star formation activity, it is expected that they were larger at higher redshift when star formation was more important on overall. This explanation holds under the assumption that the CGM responds quite quickly to the triggering or stopping of the star formation activity. Another possible explanation could be that the expansion of the universe causes a decrease of the  CGM density, preventing \MgII\ absorption  at large $b$ due to ionization.
For instance, \citet{stern_2021} explain that if we assume an isothermal density profile $ \rho(r) \propto r^{-2} \propto \rho(R_{\rm vir})\left(\frac{r}{R_{\rm vir}}\right)^{-2}$ then the cosmological evolution of $\rho(R_{\rm vir})$ implies that $\rho(r)\propto (1+z)^3\;(r/R_{\rm vir})^{-2}$. Given that the gas become self-shielded \citep{Schaye_2001} against ionizing radiation at a certain density $n_{sh}$, then the radius at which $\rho(r)$ reaches $n_{sh}$ evolves like $r_{\rm sh}\propto (1+z)^{3/2} R_{\rm vir}$. As $R_{\rm vir}\propto M_{\rm vir}^{1/3}(1+z)^{-1}$, the evolution of this self-shielding radius is $r_{\rm sh}\propto M_{\rm vir}^{1/3} (1+z)^{1/2}$, similar to the evolution of the covering fraction that we find $b_{50}\propto(1+z)^{0.8\pm0.7}$ for our sample.

Figure \ref{fig:fc_det_lim} upper right panel shows that the slope of the transition from 1 to 0 of the covering fraction is steeper for our sample than reported in previous works such as \citet{Schroetter_2021} or \citet{Dutta_2020}. The steepness of the covering fraction slope indicates how "homologous" CGMs are with each others. Indeed, a shallow covering fraction means that there are both absorptions and non absorptions cases at all $b$, meanwhile a step function would mean that the presence of absorption is completely determined by the impact parameter. In our case the steep slope could be interpreted as the fact that our strictly selected sample of isolated galaxies with $\log(M_\star/\msun)>9$ have homologous CGMs. In other word, selecting a population of isolated galaxies also select self similar CGMs.

Finally, we do not find any clear trend between the \MgII\ absorption strength and $v(2R_d)$ , $\sigma(2R_d)$ or $v/\sigma(2R_d)$. We could have expected some correlation with these kinematic parameters. Indeed, as \MgII\ absorption is often saturated, the \REW\ could have been impacted by the velocity dispersion of \MgII\ clouds along the LOS.

The different observations above lead us to a picture of a cool CGM that is anisotropic, mainly driven by the star formation activity (that is not so bursty, consistent with the relatively moderate dispersion around the MS) and with a short reconfiguration time.

Among the potential bias and limitations of this work, one can first mention the arbitrary maximum line-of-sight velocity window of $500$~km/s chosen to associate galaxies with absorption as recent simulations studies \citep{Ho_2020,Weng_2024}  indicate that absorption in this velocity window could be caused by gas outside the virial radius and/or associated with satellites. The fact that we impose no neighbour within $\pm 500$~km/s mitigate this risk of misassignment. We also underline that this bias is mostly present for low mass central galaxies. \citet{Ho_2020} estimate that for galaxies in the mass range $10^{9.5} \msun < M_\star < 10^{10} \msun$ (corresponding to our median mass of $10^{9.8} \msun$), only $\approx 15\%$ of absorption could be caused by gas outside of $R_{\rm vir}$. In our sample, we do not find any case of galaxy located at high $b$ and associated with strong absorption that could indicate such misassignement.

About the completeness, we can distinguish two different effects that could lead to some bias in our sample. First, the double galaxy detection procedure (from continuum and from [\OII] emission line) used in MEGAFLOW, could miss more low SFR galaxies at high $z$ than at low $z$. This is only true for low mass galaxies and hence do not impact the correlation with SFR shown in section \ref{sec:z,M,sfr}. Second, the isolation criterion is a bit looser at high $z$ as the completeness of low mass neighbours is lower at these redshifts.

The criterion $N_{\rm FOV} < 5$ (used to remove galaxies located near groups) depends on the field of view and, hence, evolves with redshift. However this criterion has a small effect overall as removing it from the analysis only add 4 new galaxies to the 66 of our sample.

As explained in section \ref{sec:isolation_criteria}, the isolation distance is used to define a cylinder around the LOS in which the galaxy is alone. Another neighbour galaxy could hence be present in the virial radius of the selected one and could contribute to the absorption. However this contribution is believed to be small as the neighbour would be far from the LOS.
Finally, the choice of a redshift dependent isolation distance is based on the idea that gas halos follow approximately the growth of DM halos. Our results suggest that is is probably not the case, especially at low redshifts where gas halos seems to follow the decrease of SFR. Our isolation distance may then be too conservative and could be relaxed to increase the statistic. As a test we performed the same analysis with a fixed isolation distance of 125~kpc and do not observe any qualitative change in the results.

Altogether, the potential biases from pre-selection, completeness and isolation criterion are small and they have at most a very minor influence on the result of this analysis.

\section{Conclusions}

We presented here our results for \MgII\ absortion near isolated galaxies in the MEGAFLOW sample. MEGAFLOW is a quasar absorption survey focused on the \MgII\ doublet to study the cool CGM. In the 22 quasar fields observed with both MUSE and UVES a total of 127 \MgII\ absorption systems and 1208 foreground galaxies have been identified (both from their continuum and emission lines). We focused here on a sub-sample of 66 isolated galaxies at $0.4<z<1.5$ with $\log(M_\star/\msun)>9$. We used the \galpak\ algorithm to derive the morphological and kinematic parameters for each galaxy. We then studied how \MgII\ absorption is impacted by these parameters. Our analysis on this sample led to the following conclusions:

\begin{itemize}
    \item All isolated galaxies with $\log(M_\star/\msun)>9$ having an impact parameter below 55 kpc are associated with a \MgII\ absorption with \REW\ $> 0.1$ \AA\ (Fig. \ref{fig:fc_det_lim}a).
    \item We fitted the \REW-$b$ relation and obtained results consistent with similar works on isolated galaxies such as \citet{Nielsen_2013} and \citet{Huang_2021} despite different isolation criterion (Fig. \ref{fig:fc_det_lim}a).
    \item The \MgII\ covering fraction (i.e the probability to obtain absorption at a given impact parameter) is highly dependent on the absorption strength limit (Fig. \ref{fig:fc_det_lim}b). The $50$\% probability is reached at $\approx 30$~kpc for 1\AA\ versus $\approx 65$~kpc for 0.1\AA.
    \item Isolated galaxies with large azimuthal angle ($\alpha >60$°) relative to the LOS present more extended \MgII\ absorption than galaxies with mid ($30 <\alpha <60$°) or small azimuthal angles ($\alpha <30$°) (Fig. \ref{fig:alpha_histogramms}, \ref{fig:W_b_isolated_alpha_split}).
    \item The extent, strength and probability of absorption are correlated with the SFR, consistent with the picture of galactic winds ejecting metals into the CGM to large distances (Fig. \ref{fig:multi_param_fit}).
    \item The \MgII{} halos seem to contract from high redshifts to low redshifts (Fig. \ref{fig:multi_param_fit}a) \citep[as found in][]{Schroetter_2021,Lundgren_2021} and in consequence do not grow in step with DM halos. This might be due to the overall decrease of star formation activity toward the local universe and/or to density evolution of the CGM with cosmic time.
    \item We do not observe any correlation between the \MgII\ absorption strength and the galaxy circular velocity, velocity dispersion or $v/\sigma$ (Fig. \ref{fig:2D_plots}).
\end{itemize}

Even though galaxies are often found in small groups of few members, we decided here to select isolated galaxies in order to get a sample that could be compared with idealized galaxy simulations. We hope that our results will be used in that context to better understand the underlying physics of the CGM.

\begin{acknowledgements}
This work has been carried out thanks to the support of the ANR 3DGasFlows (ANR-17-CE31-0017).
The calculations and figures have been made using the open-source softwares NUMPY \citep{Numpy_2011, Numpy}, SCIPY \citep{Scipy}, MATPLOTLIB \citep{Matplotlib}, ASTROPY \citep[][]{Astropy2013, Astropy2018} and PYMC3 \citep[][]{Hoffman_2011, Salvatier_2015}.
The data used in this work are based on observations made with ESO telescopes at the La Silla Paranal Observatory. LW and IP acknowledge funding by the European Research Council through ERC-AdG SPECMAP-CGM, GA 101020943.
\end{acknowledgements}

\section*{Data Availability}
The underlying data used for this article are available in the ESO archive (\url{http://archive.eso.org}).\\
The MEGAFLOW catalogs and reduced cubes are available at \url{https://megaflow.univ-lyon1.fr/}.\\
The catalog can be queried interactively at 
\url{https://amused.univ-lyon1.fr/megaflow/}.
The data generated for this work will be shared on reasonable request to the corresponding author.

\bibliographystyle{aa} 

\begin{appendix}

\section{Inclination and azimuthal distribution from \galpak}

\begin{figure}
	\includegraphics[width=8.5cm]{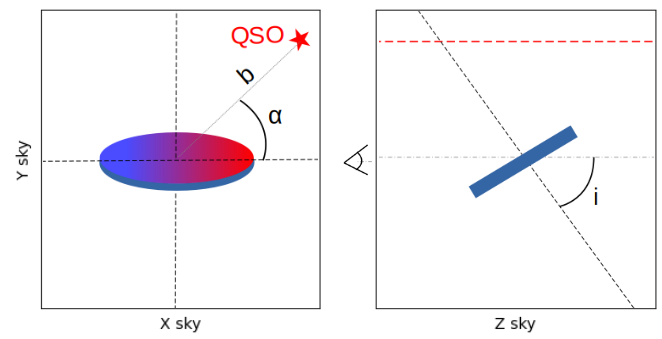} 
    \caption{Schematic representation of the azimuthal angle $\alpha$ relative to the position of the QSO and inclination angle $i$ viewed
    in the plane of the sky (along the line-of-sight) on the left (right), respectively.}
    \label{fig:sketch_angle}
\end{figure}

\begin{figure*}
	\includegraphics[width=7.5cm]{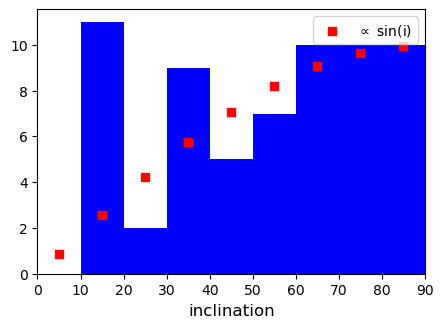} 
 	\includegraphics[width=7.5cm]{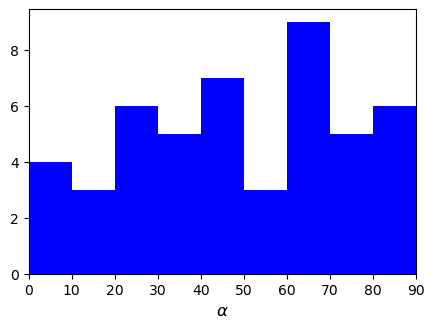} 
    \caption{\textbf{Left}: distribution of the \galpak estimated inclination for the 59 isolated rotation dominated galaxies. Red dots are proportional to the expected sinus distribution. \textbf{Right}: distribution of the azimuth angle $\alpha$ between galaxies major axis and LOS for the 46 isolated rotation dominated galaxies with sufficient inclination angle $i>30$.}
    \label{fig:inclination_distrib_isolated}
\end{figure*}

We present in Figure \ref{fig:sketch_angle} the inclination and azimuthal angle $\alpha$ between the major axis and the position of the quasar sight-line which is computed from the P.A. Figure \ref{fig:inclination_distrib_isolated}
presents the inclination $i$ and azimuthal angle $\alpha$ distribution for our sample.

\section{Mass dependence of covering fraction}
\label{sec:mass_dependence_appendix}
\begin{figure}
 	\includegraphics[width=7.5cm]{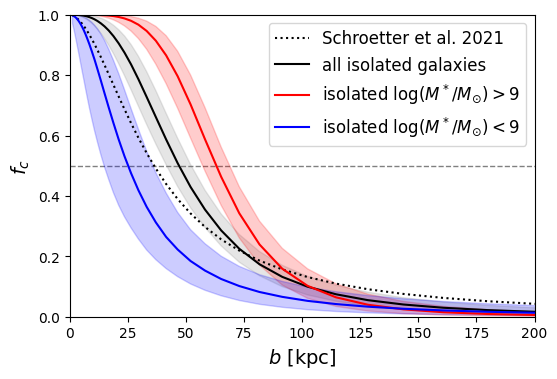} 
    \caption{Covering fraction computed for isolated galaxies with $\log(M_\star/\msun)> 9$ (our main sample, represented by the red curve), with $\log(M_\star/\msun)< 9$ (represented by the blue curve) and without mass selection (represented by the black curve). The covering fraction computed by \citet{Schroetter_2021} for MEGAFLOW primary galaxies without mass selection is represented by the black dotted line.}
\label{fig:fc_mass_dependence}
\end{figure}

In figure \ref{fig:multi_param_fit} we have shown that the \MgII\ profile and covering fraction is correlated with the SFR. The SFR is correlated with the stellar mass throught the main sequence relation. In order to visualize the mass dependence of \MgII\ absorption, we compare in Figure \ref{fig:fc_mass_dependence} the covering fraction for our isolated galaxy sample (with $\log(M_\star/\msun)> 9$) with similarly selected isolated galaxies of low mass (with $\log(M_\star/\msun)< 9$). We also show the covering fraction computed for the whole isolated galaxy sample (without mass selection) and compare it with the the covering fraction computed from \citet{Schroetter_2021} (without mass selection neither).

\section{Isolated galaxy catalog}
We present in Figure \ref{fig:catalog_1} our isolated galaxy sample. For each galaxy we show the observed velocity map obtained with \texttt{CAMEL} \citep{Epinat_2009_camel1, Epinat_2012_camel2} and the velocity map for the best \galpak model (convolved with PSF and LSF to reproduce observations). Table \ref{tab:catalog_1} also presents the main properties of the isolated galaxy sample.

\begin{figure*}
\centering
	\includegraphics[width=12cm]{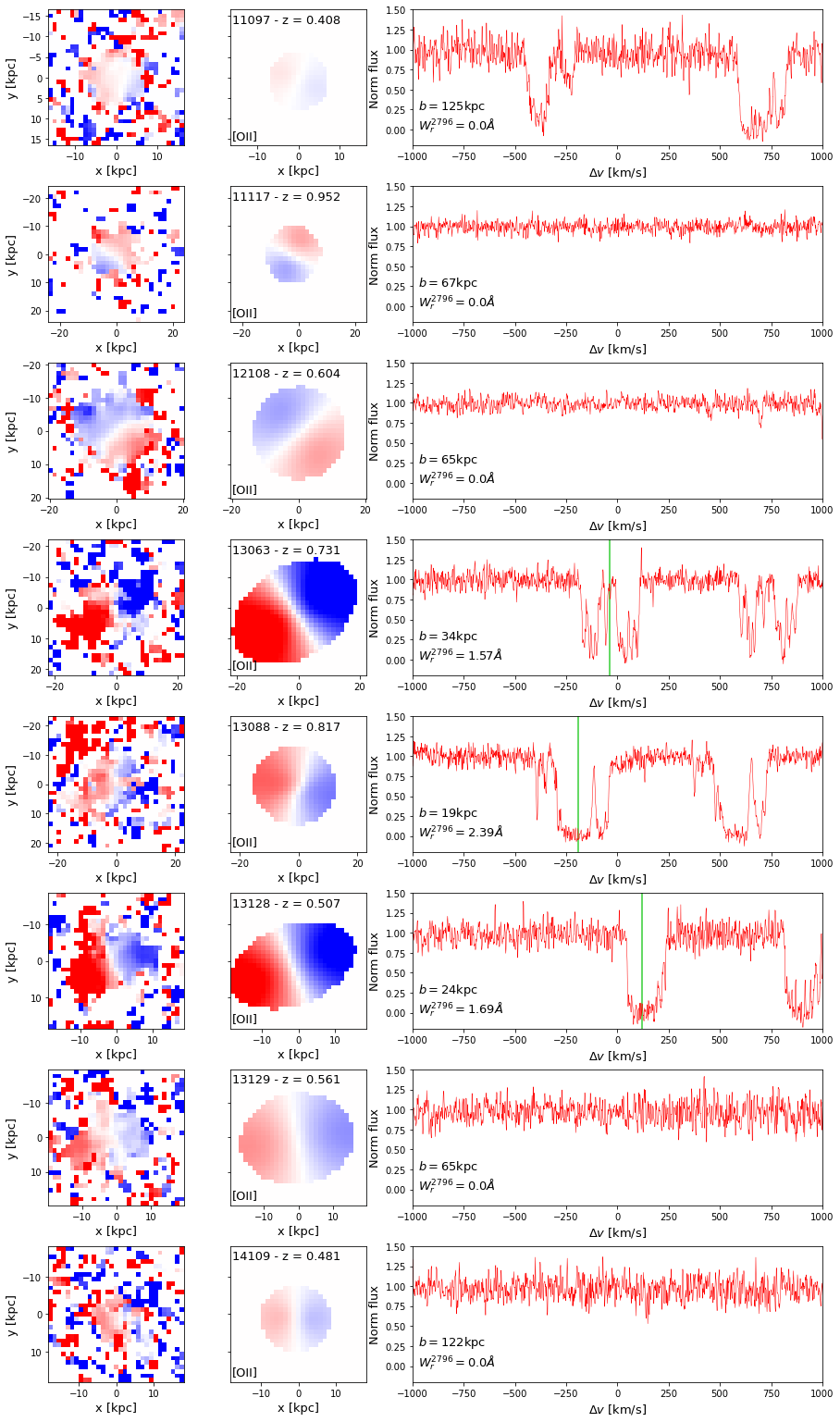} 
    \caption{Catalog of the 66 isolated galaxies. For each galaxy, the left panel presents the observed velocity map obtained with \texttt{CAMEL} with a 2 pixel spatial smoothing. The central panel presents the velocity map of the fitted \galpak model. The right panel shows the normalized UVES spectra of the quasar at the \MgII\ 2796~\AA\ doublet wavelength. The redshift of the associated absorption (if any) is indicated by a vertical green line. Galaxies 17072, 20077, 23086, 24038, 25087, 29084 have a S/N $< 4$ or are too irregular to consider the \galpak fit as reliable (they are the grey dots on Figure \ref{fig:RePSF_SNR}).}
    \label{fig:catalog_1}
\end{figure*}

\begin{figure*}
\centering
	\includegraphics[width=12cm]{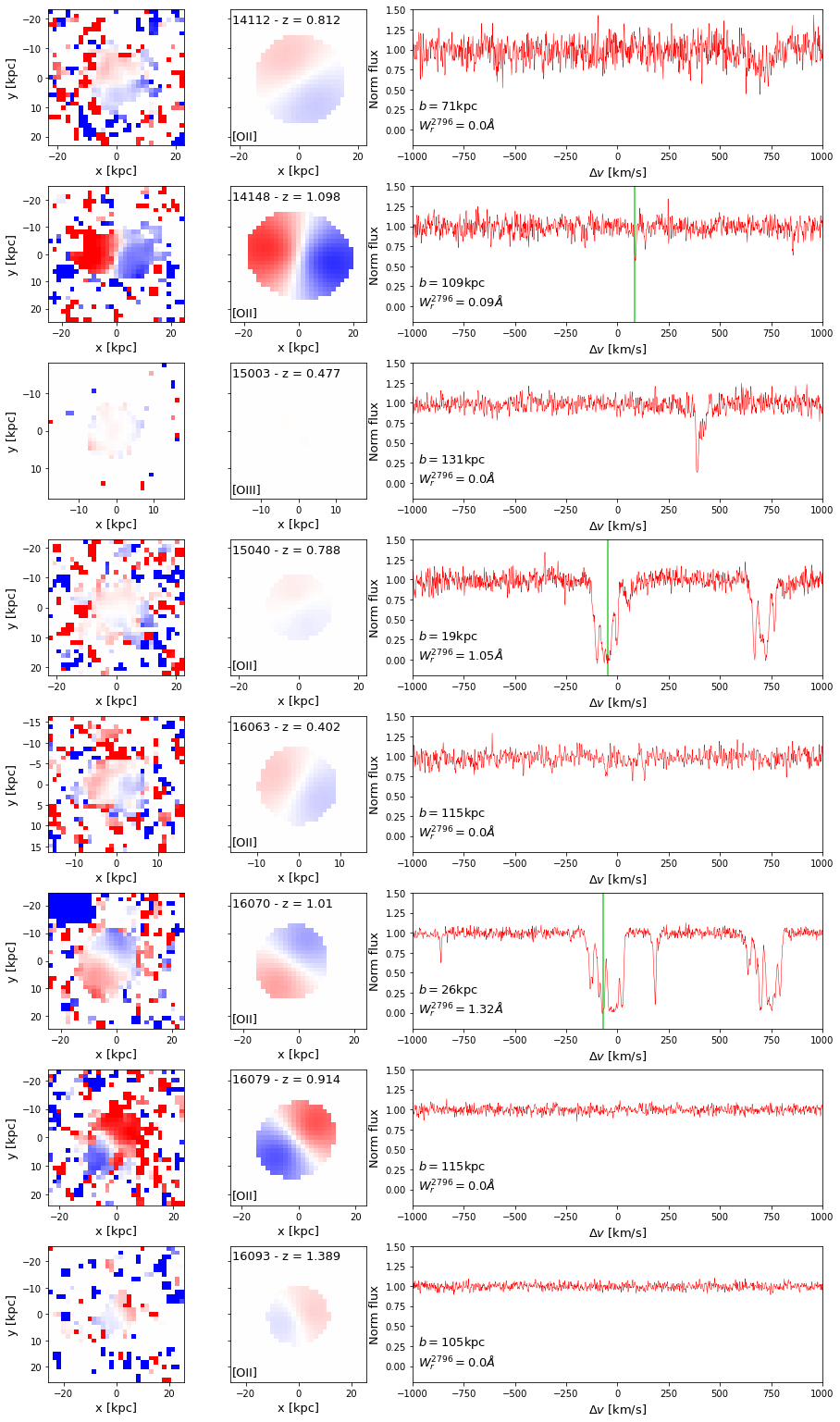} 
    \caption{Continuation of Figure \ref{fig:catalog_1}}
    \label{fig:catalog_2}
\end{figure*}

\begin{figure*}
\centering
	\includegraphics[width=12cm]{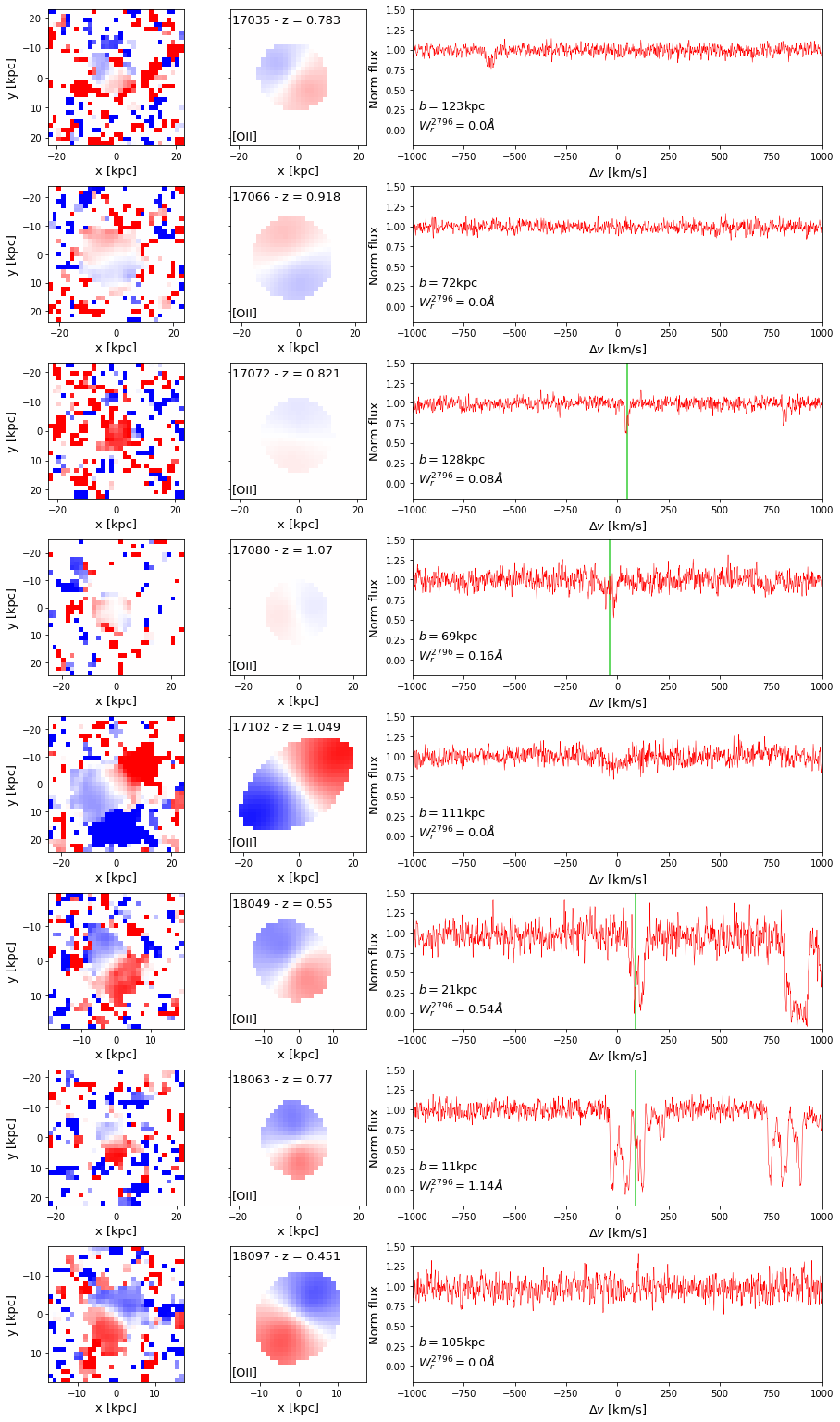} 
    \caption{Continuation of Figure \ref{fig:catalog_2}}
    \label{fig:catalog_3}
\end{figure*}

\begin{figure*}
\centering
	\includegraphics[width=12cm]{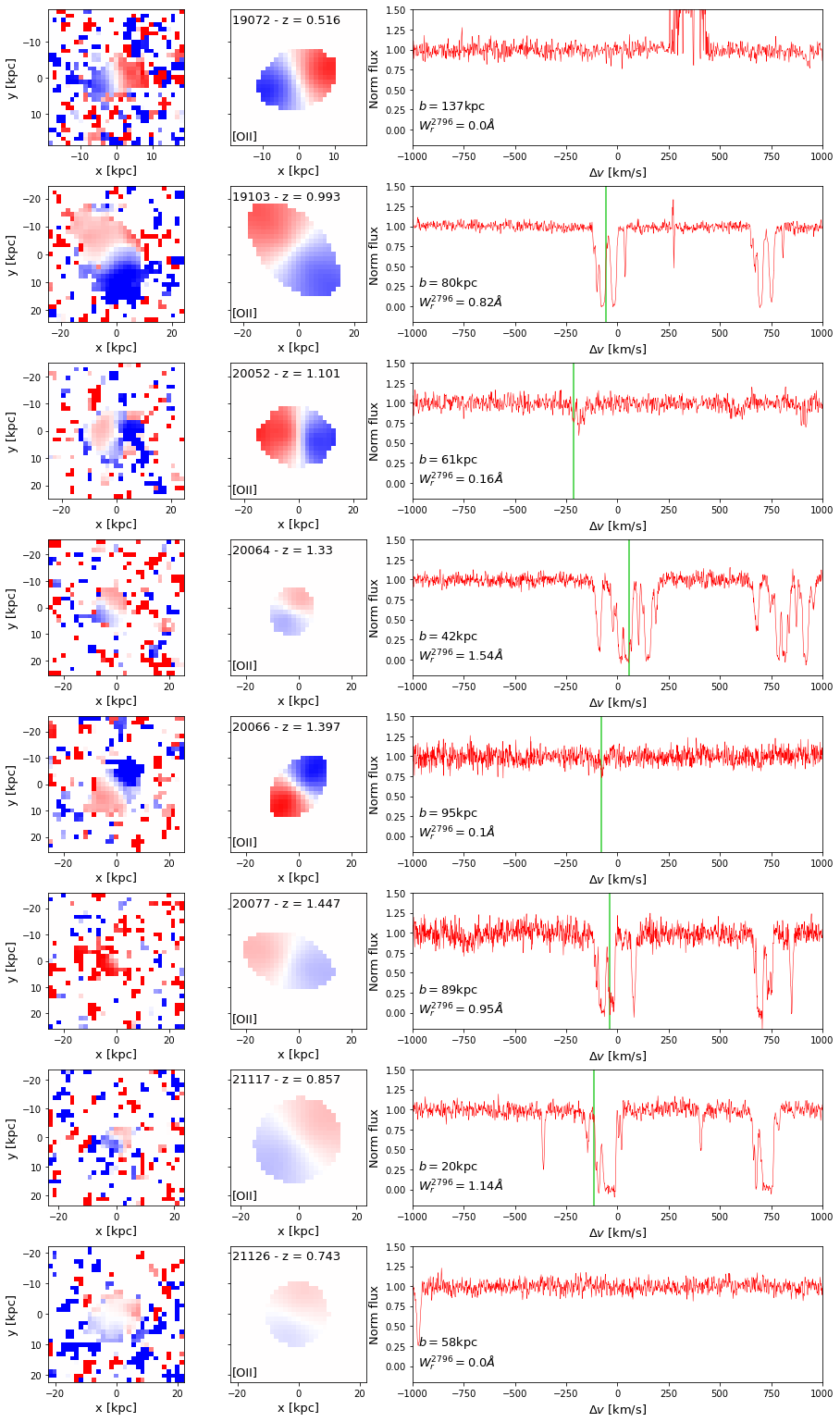} 
    \caption{Continuation of Figure \ref{fig:catalog_3}}
    \label{fig:catalog_4}
\end{figure*}

\begin{figure*}
\centering
	\includegraphics[width=12cm]{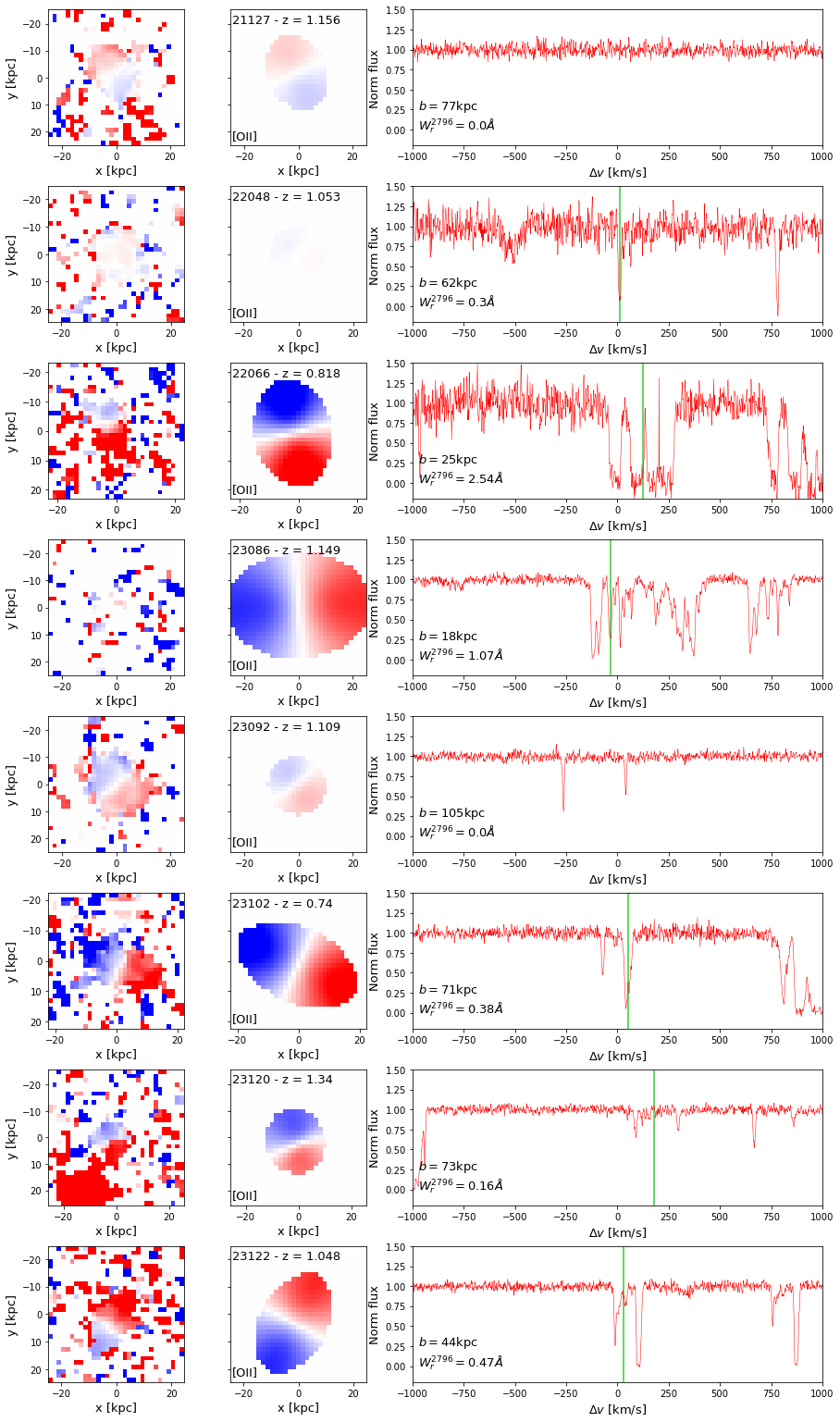} 
    \caption{Continuation of Figure \ref{fig:catalog_4}}
    \label{fig:catalog_5}
\end{figure*}

\begin{figure*}
\centering
	\includegraphics[width=12cm]{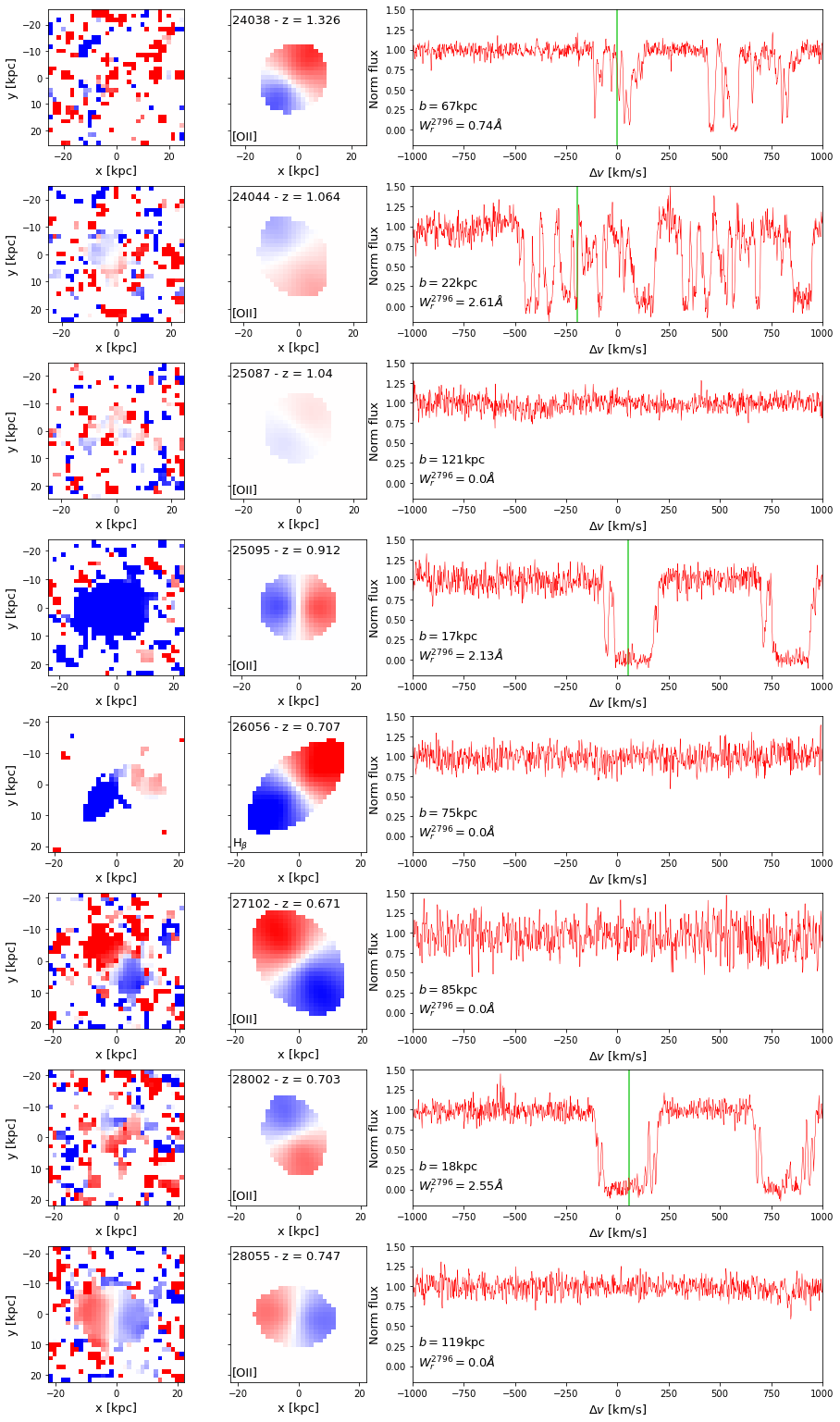} 
    \caption{Continuation of Figure \ref{fig:catalog_5}}
    \label{fig:catalog_6}
\end{figure*}

\begin{figure*}
\centering
	\includegraphics[width=12cm]{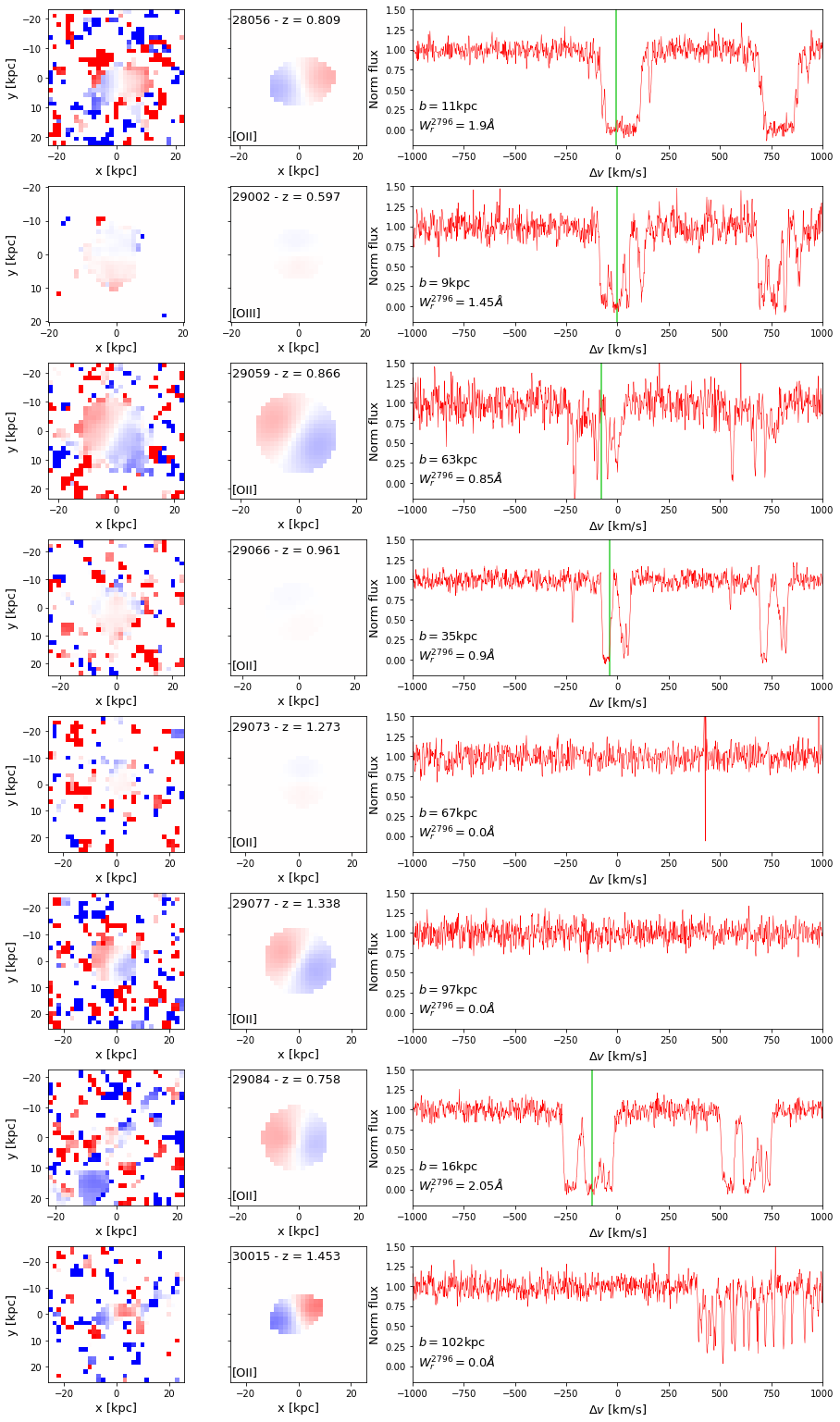} 
    \caption{Continuation of Figure \ref{fig:catalog_6}}
    \label{fig:catalog_7}
\end{figure*}

\begin{figure*}
\centering
	\includegraphics[width=12cm]{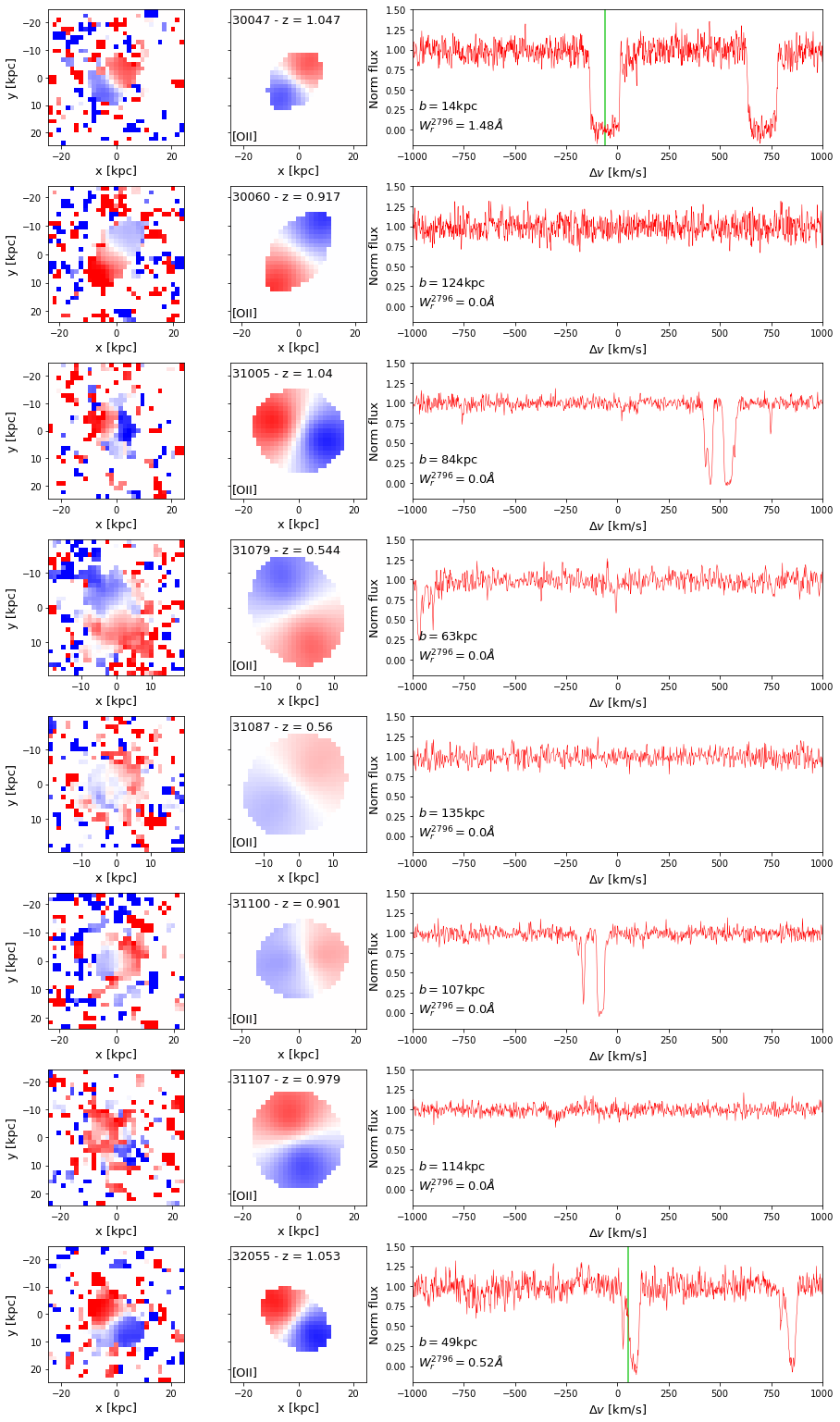} 
    \caption{Continuation of Figure \ref{fig:catalog_7}}
    \label{fig:catalog_8}
\end{figure*}

\begin{figure*}
\centering
	\includegraphics[width=12cm]{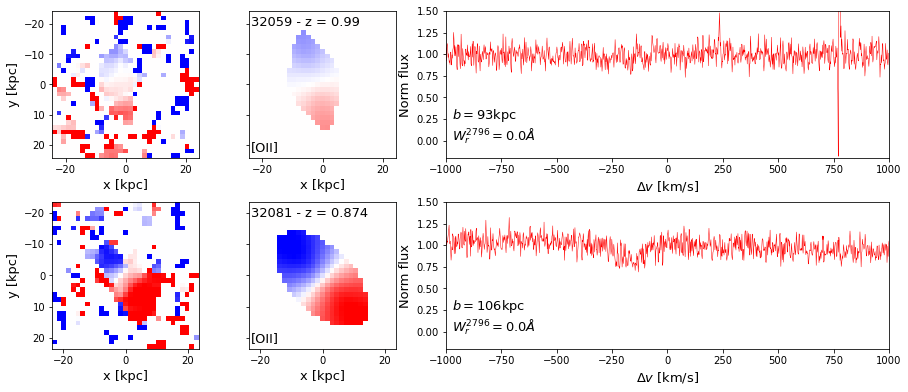} 
    \caption{Continuation of Figure \ref{fig:catalog_8}}
    \label{fig:catalog_9}
\end{figure*}

\begin{table*}
\centering
\begin{threeparttable}
\caption{Isolated galaxy sample. (1) Field name; (2) galaxy id; (3) redshift; (4) impact parameter in kpc; (5) \MgII\ rest-frame equivalent width in \AA; (6) signal to noise ratio; (7) log of the stellar mass in $\msun$, estimated from SED fitting; (8) log of the star formation rate in $\msun/$yr; (9) azimuthal angle formed with the qso sight-line, in degrees; (10) inclination in degrees; (11) circular velocity at 2$R_d$ in km/s; (12) velocity disperion at 2$R_d$ in km/s.}
\label{tab:catalog_1}
\begin{tabular}{lrllllllllllllllll}
\toprule
\hline
field & id & $z$ & $b$ & \REW & S/N & $\log(M_*)$ & log(SFR) & $\alpha$ & $i$ & $v(2R_d)$ & $\sigma(2R_d)$\\
(1) & (2) & (3) & (4) & (5) & (6) & (7) & (8) & (9) & (10) & (11) & (12)\\
\hline
\midrule
J0014m0028 & 11097 & 0.408 & 126 & $<0.08$ & 13.4 & 9.20 $\pm$ 0.01 & -0.45 $\pm$ 0.09 & 75 $\pm$ 4 & 10 $\pm$ 1 & 180 $\pm$ 22 & 29 $\pm$ 23 \\
J0014m0028 & 11117 & 0.952 & 68 & $<0.08$ & 24.9 & 9.06 $\pm$ 0.05 & 0.27 $\pm$ 0.05 & 42 $\pm$ 2 & 45 $\pm$ 2 & 120 $\pm$ 8 & 43 $\pm$ 6 \\
J0014p0912 & 12108 & 0.604 & 66 & $<0.08$ & 44.0 & 9.70 $\pm$ 0.10 & 0.82 $\pm$ 0.08 & 71 $\pm$ 2 & 17 $\pm$ 2 & 279 $\pm$ 21 & 44 $\pm$ 23 \\
J0015m0751 & 13063 & 0.731 & 35 & 1.57 & 7.8 & 10.32 $\pm$ 0.01 & 0.31 $\pm$ 0.05 & 68 $\pm$ 2 & 63 $\pm$ 3 & 276 $\pm$ 10 & 42 $\pm$ 12 \\
J0015m0751 & 13088 & 0.817 & 19 & 2.39 & 15.0 & 10.22 $\pm$ 0.03 & 0.62 $\pm$ 0.03 & 81 $\pm$ 2 & 35 $\pm$ 2 & 264 $\pm$ 22 & 55 $\pm$ 23 \\
J0015m0751 & 13128 & 0.507 & 24 & 1.69 & 17.0 & 10.29 $\pm$ 0.01 & 0.53 $\pm$ 0.03 & 74 $\pm$ 1 & 77 $\pm$ 1 & 204 $\pm$ 6 & 31 $\pm$ 7 \\
J0015m0751 & 13129 & 0.561 & 66 & $<0.08$ & 18.0 & 9.18 $\pm$ 0.08 & -0.16 $\pm$ 0.07 & 90 $\pm$ 4 & 53 $\pm$ 4 & 87 $\pm$ 7 & 20 $\pm$ 9 \\
J0058p0111 & 14109 & 0.481 & 122 & $<0.08$ & 13.7 & 10.00 $\pm$ 0.06 & -0.32 $\pm$ 0.13 & 10 $\pm$ 13 & 36 $\pm$ 10 & 140 $\pm$ 35 & 27 $\pm$ 34 \\
J0058p0111 & 14112 & 0.812 & 72 & $<0.08$ & 14.1 & 9.82 $\pm$ 0.05 & 0.11 $\pm$ 1.00 & 12 $\pm$ 7 & 17 $\pm$ 2 & 143 $\pm$ 23 & 24 $\pm$ 23 \\
J0058p0111 & 14148 & 1.098 & 109 & 0.09 & 12.7 & 10.50 $\pm$ 0.06 & 1.42 $\pm$ 0.02 & 64 $\pm$ 1 & 50 $\pm$ 2 & 192 $\pm$ 6 & 29 $\pm$ 6 \\
J0103p1332 & 15003 & 0.477 & 132 & $<0.08$ & 44.0 & 11.98 $\pm$ 0.03 & -0.23 $\pm$ 0.03 & 0 $\pm$ 5 & 16 $\pm$ 3 & 14 $\pm$ 5 & 21 $\pm$ 3 \\
J0103p1332 & 15040 & 0.788 & 20 & 1.05 & 33.7 & 9.85 $\pm$ 0.17 & 0.61 $\pm$ 0.34 & 2 $\pm$ 5 & 12 $\pm$ 2 & 152 $\pm$ 28 & 41 $\pm$ 22 \\
J0131p1303 & 16063 & 0.402 & 115 & $<0.08$ & 29.1 & 9.50 $\pm$ 0.15 & -0.20 $\pm$ 0.06 & 11 $\pm$ 7 & 30 $\pm$ 9 & 125 $\pm$ 32 & 24 $\pm$ 29 \\
J0131p1303 & 16070 & 1.010 & 26 & 1.31 & 45.0 & 9.67 $\pm$ 0.04 & 0.98 $\pm$ 0.06 & 72 $\pm$ 1 & 55 $\pm$ 2 & 102 $\pm$ 4 & 35 $\pm$ 3 \\
J0131p1303 & 16079 & 0.914 & 115 & $<0.08$ & 7.5 & 9.87 $\pm$ 0.08 & 0.76 $\pm$ 1.00 & 2 $\pm$ 2 & 61 $\pm$ 3 & 177 $\pm$ 8 & 33 $\pm$ 10 \\
J0131p1303 & 16093 & 1.389 & 105 & $<0.08$ & 14.5 & 9.32 $\pm$ 0.17 & 0.36 $\pm$ 0.14 & 24 $\pm$ 6 & 73 $\pm$ 5 & 48 $\pm$ 5 & 7 $\pm$ 6 \\
J0134p0051 & 17035 & 0.783 & 123 & $<0.08$ & 13.1 & 9.46 $\pm$ 0.14 & -0.38 $\pm$ 0.12 & 61 $\pm$ 6 & 81 $\pm$ 7 & 84 $\pm$ 7 & 19 $\pm$ 11 \\
J0134p0051 & 17066 & 0.918 & 72 & $<0.08$ & 36.0 & 10.33 $\pm$ 0.12 & 0.98 $\pm$ 0.01 & 12 $\pm$ 3 & 20 $\pm$ 2 & 162 $\pm$ 14 & 25 $\pm$ 15 \\
J0134p0051 & 17072 & 0.821 & 129 & 0.08 & 2.5 & 10.64 $\pm$ 0.06 & 0.40 $\pm$ 0.12 & 84 $\pm$ 90 & 23 $\pm$ 23 & 79 $\pm$ 68 & 64 $\pm$ 29 \\
J0134p0051 & 17080 & 1.070 & 69 & 0.16 & 18.0 & 9.55 $\pm$ 0.10 & 0.61 $\pm$ 0.10 & 79 $\pm$ 15 & 13 $\pm$ 12 & 141 $\pm$ 76 & 61 $\pm$ 33 \\
J0134p0051 & 17102 & 1.049 & 112 & $<0.08$ & 14.9 & 9.58 $\pm$ 0.05 & 0.93 $\pm$ 0.13 & 77 $\pm$ 0 & 84 $\pm$ 2 & 149 $\pm$ 3 & 33 $\pm$ 6 \\
J0145p1056 & 18049 & 0.550 & 21 & 0.54 & 19.8 & 9.51 $\pm$ 0.03 & -0.19 $\pm$ 0.09 & 10 $\pm$ 1 & 57 $\pm$ 0 & 88 $\pm$ 7 & 24 $\pm$ 6 \\
J0145p1056 & 18063 & 0.770 & 12 & 1.14 & 8.0 & 9.59 $\pm$ 0.11 & -0.25 $\pm$ 0.29 & 89 $\pm$ 4 & 86 $\pm$ 3 & 116 $\pm$ 9 & 21 $\pm$ 12 \\
J0145p1056 & 18097 & 0.451 & 106 & $<0.08$ & 19.8 & 10.07 $\pm$ 0.01 & 0.37 $\pm$ 0.05 & 55 $\pm$ 1 & 69 $\pm$ 2 & 144 $\pm$ 5 & 24 $\pm$ 6 \\
J0800p1849 & 19072 & 0.516 & 138 & $<0.08$ & 14.4 & 9.76 $\pm$ 0.04 & 0.02 $\pm$ 0.39 & 48 $\pm$ 2 & 69 $\pm$ 2 & 148 $\pm$ 9 & 25 $\pm$ 11 \\
J0800p1849 & 19103 & 0.993 & 80 & 0.82 & 25.9 & 10.18 $\pm$ 0.03 & 1.38 $\pm$ 0.02 & 62 $\pm$ 1 & 70 $\pm$ 1 & 104 $\pm$ 4 & 50 $\pm$ 3 \\
J0838p0257 & 20052 & 1.101 & 62 & 0.16 & 32.3 & 10.39 $\pm$ 0.09 & 1.35 $\pm$ 0.01 & 83 $\pm$ 0 & 47 $\pm$ 0 & 137 $\pm$ 6 & 42 $\pm$ 4 \\
J0838p0257 & 20064 & 1.330 & 42 & 1.54 & 11.9 & 9.63 $\pm$ 0.11 & 0.63 $\pm$ 0.19 & 48 $\pm$ 1 & 51 $\pm$ 1 & 80 $\pm$ 7 & 35 $\pm$ 7 \\
J0838p0257 & 20066 & 1.397 & 96 & 0.10 & 19.9 & 9.92 $\pm$ 0.10 & 1.30 $\pm$ 0.15 & 62 $\pm$ 1 & 69 $\pm$ 1 & 190 $\pm$ 4 & 43 $\pm$ 6 \\
J0838p0257 & 20077 & 1.447 & 89 & 0.95 & 3.8 & 11.29 $\pm$ 0.24 & 1.38 $\pm$ 0.05 & 17 $\pm$ 4 & 68 $\pm$ 4 & 48 $\pm$ 21 & 90 $\pm$ 14 \\
J0937p0656 & 21117 & 0.857 & 20 & 1.15 & 7.2 & 10.41 $\pm$ 0.01 & -0.09 $\pm$ 0.04 & 50 $\pm$ 62 & 36 $\pm$ 21 & 90 $\pm$ 74 & 95 $\pm$ 26 \\
J0937p0656 & 21126 & 0.743 & 59 & $<0.08$ & 26.7 & 9.05 $\pm$ 0.11 & -0.15 $\pm$ 0.04 & 9 $\pm$ 10 & 24 $\pm$ 14 & 139 $\pm$ 43 & 23 $\pm$ 42 \\
J0937p0656 & 21127 & 1.156 & 78 & $<0.08$ & 34.3 & 9.39 $\pm$ 0.07 & 0.40 $\pm$ 0.04 & 5 $\pm$ 1 & 85 $\pm$ 3 & 41 $\pm$ 2 & 36 $\pm$ 2 \\
J1039p0714 & 22048 & 1.053 & 62 & 0.30 & 41.2 & 9.12 $\pm$ 0.10 & 0.70 $\pm$ 0.03 & 46 $\pm$ 3 & 78 $\pm$ 3 & 13 $\pm$ 3 & 44 $\pm$ 1 \\
J1039p0714 & 22066 & 0.818 & 26 & 2.54 & 5.2 & 10.25 $\pm$ 0.07 & 0.41 $\pm$ 0.15 & 63 $\pm$ 2 & 66 $\pm$ 3 & 252 $\pm$ 12 & 42 $\pm$ 16 \\
J1107p1021 & 23086 & 1.149 & 19 & 1.07 & 1.6 & 11.50 $\pm$ 0.05 & 0.27 $\pm$ 0.06 & 13 $\pm$ 90 & 48 $\pm$ 22 & 178 $\pm$ 83 & 94 $\pm$ 67 \\
J1107p1021 & 23092 & 1.109 & 105 & $<0.08$ & 27.2 & 9.60 $\pm$ 0.03 & 1.50 $\pm$ 0.07 & 37 $\pm$ 1 & 34 $\pm$ 1 & 136 $\pm$ 6 & 56 $\pm$ 4 \\
J1107p1021 & 23102 & 0.740 & 72 & 0.38 & 8.9 & 9.43 $\pm$ 0.06 & 0.00 $\pm$ 0.15 & 73 $\pm$ 2 & 70 $\pm$ 3 & 188 $\pm$ 8 & 35 $\pm$ 12 \\
J1107p1021 & 23120 & 1.340 & 73 & 0.16 & 7.6 & 9.95 $\pm$ 0.15 & 0.76 $\pm$ 0.32 & 38 $\pm$ 3 & 42 $\pm$ 2 & 201 $\pm$ 13 & 41 $\pm$ 19 \\
J1107p1021 & 23122 & 1.048 & 44 & 0.47 & 5.0 & 9.77 $\pm$ 0.12 & 0.59 $\pm$ 1.00 & 23 $\pm$ 2 & 75 $\pm$ 3 & 145 $\pm$ 5 & 23 $\pm$ 7 \\
J1107p1757 & 24038 & 1.326 & 67 & 0.73 & 1.5 & 10.42 $\pm$ 0.24 & 1.15 $\pm$ 0.32 & 59 $\pm$ 32 & 68 $\pm$ 18 & 234 $\pm$ 64 & 76 $\pm$ 67 \\
J1107p1757 & 24044 & 1.064 & 23 & 2.61 & 13.1 & 9.61 $\pm$ 0.24 & 0.46 $\pm$ 0.16 & 66 $\pm$ 8 & 80 $\pm$ 8 & 70 $\pm$ 20 & 46 $\pm$ 16 \\
J1236p0725 & 25087 & 1.040 & 122 & $<0.08$ & 5.6 & 9.49 $\pm$ 0.26 & -0.11 $\pm$ 0.28 & 47 $\pm$ 32 & 42 $\pm$ 20 & 42 $\pm$ 28 & 9 $\pm$ 26 \\
J1236p0725 & 25095 & 0.912 & 18 & 2.13 & 10.4 & 10.09 $\pm$ 0.01 & 0.46 $\pm$ 0.14 & 22 $\pm$ 1 & 57 $\pm$ 1 & 191 $\pm$ 4 & 31 $\pm$ 5 \\
J1314p0657 & 26056 & 0.707 & 76 & $<0.08$ & 12.4 & 10.42 $\pm$ 0.01 & 1.11 $\pm$ 0.02 & 26 $\pm$ 2 & 75 $\pm$ 2 & 198 $\pm$ 6 & 30 $\pm$ 7 \\
J1352p0614 & 27102 & 0.671 & 86 & $<0.08$ & 8.6 & 10.09 $\pm$ 0.03 & 0.28 $\pm$ 0.12 & 43 $\pm$ 1 & 80 $\pm$ 3 & 185 $\pm$ 6 & 28 $\pm$ 6 \\
J1358p1145 & 28002 & 0.703 & 18 & 2.55 & 8.8 & 9.73 $\pm$ 0.04 & -0.46 $\pm$ 0.44 & 83 $\pm$ 7 & 59 $\pm$ 11 & 132 $\pm$ 20 & 27 $\pm$ 23 \\
J1358p1145 & 28055 & 0.747 & 120 & $<0.08$ & 46.9 & 9.33 $\pm$ 0.04 & 0.58 $\pm$ 0.03 & 60 $\pm$ 1 & 63 $\pm$ 1 & 110 $\pm$ 2 & 29 $\pm$ 2 \\
J1358p1145 & 28056 & 0.809 & 11 & 1.90 & 31.9 & 9.47 $\pm$ 0.03 & 0.27 $\pm$ 0.05 & 79 $\pm$ 2 & 70 $\pm$ 2 & 53 $\pm$ 4 & 48 $\pm$ 3 \\
\hline
\bottomrule
\end{tabular}
\end{threeparttable}
\end{table*}

\begin{table*}
\centering
\begin{threeparttable}
\caption{Continuation of table \ref{tab:catalog_1}}
\label{tab:catalog_2}
\begin{tabular}{lrllllllllllllllll}
\toprule
\hline
field & id & $z$ & $b$ & \REW & S/N & $\log(M_*/\msun)$ & log(SFR/($\msun$/yr)) & $\alpha$ & $i$ & $v(2R_d)$ & $\sigma(2R_d)$\\
(1) & (2) & (3) & (4) & (5) & (6) & (7) & (8) & (9) & (10) & (11) & (12)\\
\hline
\midrule
J1425p1209 & 29002 & 0.597 & 9 & 1.45 & 30.4 & 10.21 $\pm$ 0.01 & 0.93 $\pm$ 0.05 & 84 $\pm$ 2 & 16 $\pm$ 1 & 58 $\pm$ 6 & 34 $\pm$ 2 \\
J1425p1209 & 29059 & 0.866 & 64 & 0.85 & 52.0 & 9.49 $\pm$ 0.03 & 0.65 $\pm$ 0.03 & 66 $\pm$ 1 & 39 $\pm$ 2 & 90 $\pm$ 4 & 26 $\pm$ 3 \\
J1425p1209 & 29066 & 0.961 & 35 & 0.90 & 34.9 & 9.25 $\pm$ 0.08 & 0.51 $\pm$ 0.04 & 12 $\pm$ 4 & 12 $\pm$ 2 & 36 $\pm$ 18 & 48 $\pm$ 3 \\
J1425p1209 & 29073 & 1.273 & 67 & $<0.08$ & 18.2 & 9.69 $\pm$ 0.32 & 0.42 $\pm$ 0.16 & 61 $\pm$ 90 & 31 $\pm$ 29 & 25 $\pm$ 17 & 43 $\pm$ 4 \\
J1425p1209 & 29077 & 1.338 & 97 & $<0.08$ & 18.1 & 9.75 $\pm$ 0.36 & 0.54 $\pm$ 0.28 & 87 $\pm$ 5 & 33 $\pm$ 7 & 115 $\pm$ 26 & 24 $\pm$ 24 \\
J1425p1209 & 29084 & 0.758 & 17 & 2.05 & 5.6 & 10.35 $\pm$ 0.01 & -0.20 $\pm$ 0.06 & 70 $\pm$ 33 & 34 $\pm$ 21 & 145 $\pm$ 71 & 50 $\pm$ 52 \\
J1509p1506 & 30015 & 1.453 & 103 & $<0.08$ & 6.5 & 10.77 $\pm$ 0.09 & 0.75 $\pm$ 0.02 & 66 $\pm$ 8 & 78 $\pm$ 6 & 94 $\pm$ 10 & 17 $\pm$ 12 \\
J1509p1506 & 30047 & 1.047 & 15 & 1.48 & 22.5 & 9.10 $\pm$ 0.04 & 0.62 $\pm$ 0.06 & 28 $\pm$ 1 & 81 $\pm$ 3 & 125 $\pm$ 7 & 20 $\pm$ 8 \\
J1509p1506 & 30060 & 0.917 & 124 & $<0.08$ & 30.3 & 9.46 $\pm$ 0.10 & 0.49 $\pm$ 0.06 & 36 $\pm$ 1 & 82 $\pm$ 1 & 112 $\pm$ 3 & 30 $\pm$ 3 \\
J2137p0012 & 31005 & 1.040 & 85 & $<0.08$ & 8.8 & 9.69 $\pm$ 0.01 & 0.50 $\pm$ 0.45 & 30 $\pm$ 6 & 61 $\pm$ 9 & 228 $\pm$ 28 & 45 $\pm$ 31 \\
J2137p0012 & 31079 & 0.544 & 63 & $<0.08$ & 24.3 & 9.76 $\pm$ 0.01 & 0.06 $\pm$ 0.23 & 19 $\pm$ 1 & 85 $\pm$ 3 & 115 $\pm$ 4 & 30 $\pm$ 7 \\
J2137p0012 & 31087 & 0.560 & 135 & $<0.08$ & 12.9 & 10.83 $\pm$ 0.01 & 0.70 $\pm$ 0.01 & 13 $\pm$ 6 & 56 $\pm$ 8 & 78 $\pm$ 18 & 42 $\pm$ 15 \\
J2137p0012 & 31100 & 0.901 & 107 & $<0.08$ & 9.8 & 9.13 $\pm$ 0.19 & -0.33 $\pm$ 0.12 & 45 $\pm$ 4 & 86 $\pm$ 3 & 88 $\pm$ 5 & 15 $\pm$ 7 \\
J2137p0012 & 31107 & 0.979 & 114 & $<0.08$ & 8.8 & 10.36 $\pm$ 0.04 & 0.84 $\pm$ 0.03 & 37 $\pm$ 4 & 49 $\pm$ 5 & 208 $\pm$ 18 & 47 $\pm$ 20 \\
J2152p0625 & 32055 & 1.053 & 50 & 0.52 & 9.1 & 10.09 $\pm$ 0.05 & 1.27 $\pm$ 0.10 & 4 $\pm$ 1 & 73 $\pm$ 1 & 166 $\pm$ 4 & 25 $\pm$ 5 \\
J2152p0625 & 32059 & 0.990 & 94 & $<0.08$ & 19.8 & 9.22 $\pm$ 0.24 & 0.16 $\pm$ 0.06 & 30 $\pm$ 3 & 82 $\pm$ 3 & 52 $\pm$ 6 & 33 $\pm$ 5 \\
J2152p0625 & 32081 & 0.874 & 106 & $<0.08$ & 7.1 & 9.83 $\pm$ 0.09 & 0.51 $\pm$ 1.00 & 57 $\pm$ 1 & 77 $\pm$ 2 & 179 $\pm$ 5 & 27 $\pm$ 5 \\
\hline
\bottomrule
\end{tabular}
\end{threeparttable}
\end{table*}

\end{appendix}

%
%

\end{document}